\documentclass[11pt]{article}
\usepackage[top=1in, bottom=1in, left=1in, right=1in]{geometry}
\usepackage{setspace}

\usepackage[T1]{fontenc}
\usepackage{times}
\usepackage{booktabs}
\usepackage{rotating}
\usepackage{graphicx}
\usepackage[section]{placeins} 
\usepackage[large, bf]{caption}
\usepackage{subcaption}
\usepackage{threeparttable}
\usepackage{pdfpages}
\usepackage{palatino}
\usepackage{pdflscape}
\usepackage{textcomp}
\usepackage{longtable}
\usepackage{nicefrac}
\usepackage{adjustbox}	
\usepackage[hyphens]{url}

\usepackage{natbib}
\bibpunct{(}{)}{;}{a}{,}{,}

\usepackage{amsmath, amsfonts, amssymb, amsthm}

\usepackage{mathpazo} 
\usepackage[pdftex]{hyperref}
\hypersetup{colorlinks, citecolor=blue, filecolor=blue, linkcolor=blue, urlcolor=blue}

\def\urltilda{\kern -.15em\lower .7ex\hbox{\~{}}\kern .04em}

\usepackage{titlesec}

\titlespacing*{\section}{0pt}{1.5ex plus 1ex minus .2ex}{0.8ex plus .2ex}
\titlespacing*{\subsection}{0pt}{1.2ex plus 1ex minus .2ex}{0.8ex plus .2ex}

\usepackage{dcolumn}
\newcolumntype{d}[0]{D{.}{.}{5}}
\usepackage{mathtools}
\usepackage{tikz}
\usepackage{graphicx}
\usepackage[space]{grffile}

\newtheorem*{rem*}{\protect\remarkname}
\providecommand{\remarkname}{Remark}

\usepackage[official]{eurosym}
\usetikzlibrary{arrows,positioning} 
\usetikzlibrary{decorations.pathreplacing}
\usetikzlibrary{decorations.pathmorphing}
\usetikzlibrary{intersections}
\usepackage{float}
\usepackage[american]{babel}

\usepackage{tikz}
\usetikzlibrary{patterns}
\usepackage{adjustbox}
\usepackage{hyperref}

\usepackage{booktabs}
\usepackage{graphicx}
\usepackage{cleveref}
\usepackage{stfloats}
\usepackage{tabulary}
\usepackage{bbm} 
\usepackage{siunitx}

\begin{document}

\title{\textbf{\LARGE{The Impact of Shared Telecom Infrastructure on Digital Connectivity and Inclusion}} 
\thanks{This research benefited from the financial support of the FIT-IN program hosted by the Toulouse School of Economics and was presented at the conference organized by that program in December 2022, as well as the TPRC 52 Conference. The Authors are grateful for data support from Saniya Ansar (World Bank) and Natalia Veligura (IFC), to Abdullah L. Alnasser and Shangrong Chen for their research assistance, and for comments from conference participants as well as Carlo Maria Rossotto and Tania P. Begazo Gomez.}\ \vspace{0.2cm} \normalsize
}

\author{%
Georges\,V.~Houngbonon\thanks{International Finance Corporation (IFC) and GWU Competition and Innovation Lab.  
\textit{}: \texttt{ghoungbonon@ifc.org}.}
\and
Marc~Ivaldi\thanks{Toulouse School of Economics (TSE).  
\textit{}: \texttt{marc.ivaldi@tse-fr.eu}.}
\and
Emil~Palikot\thanks{Stanford University.  
\textit{}: \texttt{palikot@stanford.edu}.}
\and
Davide~Strusani\thanks{International Finance Corporation (IFC).  
\textit{}: \texttt{dstrusani@ifc.org}.}
}

\date{\today}

\maketitle

\thispagestyle{empty} 
\setcounter{page}{0}

\begin{abstract}\noindent
Nearly half the world remains offline, and capital scarcity stalls new network buildouts. Sharing existing mobile towers could accelerate connectivity. We assemble data on 107 tower-sharing deals in 28 low-income countries (2008–20) and estimate staggered difference-in-differences effects. Two years after a transaction covering over 1,000 towers, the PPP-adjusted mobile-price index  falls \$1.60 (s.e.1.10) from a baseline of \$3.16, while data prices drop \$1.00 (0.29), baseline \$3.41/GB. The number of mobile connections increases. Rural internet access increases by 4.7 pp and female-headed households by 3.6 pp. Tower-sharing agreements increase product market competition as measured by Herfindahl–Hirschman Index.

\end{abstract}



\newpage{}

\section{Introduction}

Digital networks are widely credited with boosting productivity, catalysing firm entry, and expanding access to education and health services \citep{Czernich2011,HjortPoulsen2019,MalamudPop2011}. Yet, after two decades of mobile diffusion, one person in three remains offline, and many more connect only through low-quality links \citep{ITU2023Facts,itugcr2022}. This connectivity gap broadly results from a number of challenges, including limited affordability, investment, and low literacy and income \citep{chen2021}, as well as limited availability of relevant content and access to electricity \citep{armey2016, houngbonon2020, houngbonon2021}.

In the mobile telecommunications industry, infrastructure-based competition, whereby mobile operators deploy their own network infrastructure and compete for end-users, has been considered by regulators to improve service affordability and boost investment \citep{houngbonon2016, jeanjean2017, genakos2018}. However, such a policy met with several challenges, especially the need for frequent investment in network upgrade due to fast-paced technological progress \citep{koh2006functional}. In developing countries, this challenge is compounded with low ability to pay for connectivity services by end-users. This has resulted in the development of shared infrastructure business models \citep{ds_gvh_2020}, both bilateral agreements between network operators \citep{koutroumpis2021} or multilateral agreements through independent infrastructure operators \citep{gvh_etal_2021}.

Independent tower companies (\emph{towercos}) offer an alternative.  By taking over the passive parts of mobile sites—towers, masts, power supply—then leasing space on those structures to \emph{any} operator, including the original seller, on equal terms. Turning towers into shared “rental property’’ lets operators avoid building parallel grids, frees cash for upgrading radio spectrum, and allows new or smaller carriers to enter without heavy upfront capex. 

This paper provides new causal evidence. We compile a global database that canvasses tower-sharing activity in 182 developing countries and identifies 107 transactions completed between 2008 and 2020 in 28 different countries. These deals are matched to annual series on retail mobile tariffs, mobile- and Internet-connectivity outcomes, macro controls, and basic demographics. Because tower sales close in different years across countries, we exploit the staggered timing with the interaction-weighted estimator of \citet{sun2021estimating}. Our baseline definition of treatment requires a deal to involve at least 1000 towers—a threshold widely viewed as the minimum for stand-alone profitability. \footnote{We show that the main results are robust to changes in the cutoff value.}

We find that tower-sharing transactions reduce prices and expand connectivity. The PPP-adjusted mobile‐price index, which averages \$3.16 in 2010, declines by \$1.60 (s.e. \$1.10); the data-price index, baseline \$3.41 per GB, falls by \$1.00 (s.e.\ \$0.29). We measure uptake by the number of connections (in thousands) and the number of mobile broadband-capable connections. We estimate that the number of mobile connections increases by 1422 thousand (s.e. 385); the impact on the mobile broadband-capable is high but statistically insignificant at 600 (s.e. 477) . The share of rural households with Internet access increases by 4.7 percentage points (s.e.\ 1.5 pp), and the share of women-headed households by 3.6 percentage points (s.e.\ 1.2 pp). Retail market concentration, measured by the Herfindahl–Hirschman Index, drops by 5.3 percent (s.e.\ 2.0 percent). These estimates are robust to alternative tower-count thresholds and additional control variables, and event-study coefficients show no significant pre-treatment trends.

We evaluate heterogeneity of treatment effects and find that the impact of tower-sharing depends on contract form and, to a lesser extent, on deal size. Sale-and-leaseback transactions—where the selling operator remains an anchor tenant—produce the largest price reductions, while extending a deal beyond roughly 2 000 towers adds little additional effect. These patterns underscore that contractual design, rather than scale alone, governs the efficacy of tower sharing and suggest that well-structured tower-company arrangements can work alongside spectrum- and service-based instruments to narrow the digital divide.

The findings of this paper fit into the literature on the welfare effects of market structure in the mobile industry. Most studies, including \cite{genakos2018}, \cite{jeanjean2017} and \citet{elliott2021} investigated the impact of infrastructure-based competition on price and investment. In this study, we focus on shared telecom infrastructure as an alternative to infrastructure-based competition. Other studies like \citet{koutroumpis2021} evaluate the effects of shared telecom infrastructure, but focus on bilateral sharing, whereas this paper considers multilateral infrastructure sharing, a growing trend in the telecom sector.

The remainder of the paper is organized as follows. Section \ref{sect:background} provides some background on the digital divide and the business models of shared telecom infrastructure. Section \ref{sect:literature} provides an overview of the related literature. Section \ref{sect:concept} presents the conceptual framework, while Section \ref{sect:data} presents the data with descriptive statistics. Section \ref{sect:estimation-results} presents the econometric models, the estimation strategies and reports the results. Section \ref{sect:conclusion} concludes.

\section{Background on the digital divide and shared telecom infrastructure}
\label{sect:background}

$\blacksquare$ \textbf{The digital divide}

The digital divide is defined as a \textit{division between people who have access to and use of digital technologies and those who do not} \citep{vandijk2020}. As such, it involves (i) an access dimension which pertains to the availability and uptake of digital devices, connectivity services, or applications; and (ii) a usage dimension that relates to the usage intensity of the digital technologies, including the usage capabilities (e.g., literacy and skills) and the ability to direct access towards productive usage.

Several studies are available on the access dimension. Most studies focus on connectivity devices and services \citep{itugcr2022, gsma2022}, but a growing set of studies document the applications layer of the access dimension, especially e-commerce and digital financial services \citep{findex2022}. As of 2021, the global digital divide in access remained significant: nearly 3 billion people remain unconnected, 90 percent of which live in low or middle-income countries, especially in India (787mn), Pakistan (166mn), Nigeria (133mn), Indonesia (127mn) and Bangladesh (124mn).\footnote{See \cite{itugcr2022}.} The unconnected are largely poorer, less educated, female, persons with disabilities, and rural, with 234 million fewer women than men using mobile Internet in developing countries \citep{gsma_inclusion2022}.

$\blacksquare$ \textbf{Shared telecom infrastructure}

Telecommunications network infrastructure includes facilities such as fiber optic cables, towers, ducts, poles, submarine landing stations, data centers and cabinets, as well as resources such as radio frequency spectrum and energy. These infrastructure can be shared among telecom service providers (e.g., mobile network operators and Internet service providers) under a variety of business models which can be grouped according to the degree of ownership of the infrastructure by service providers \citep{gallegos2018, ds_gvh_2020}.\footnote{Telecom infrastructure can also rely on infrastructure from other sectors such as railways, oil and gas pipeline, electric distribution systems, and city infrastructure like sewage systems, newsstands and bus stations.}

Under full ownership, the network infrastructure can be shared among service providers through bilateral agreements or access regulation \citep{koutroumpis2021}. Bilateral agreements involve commercial contracts between two telecom service providers, of which at least one own a network infrastructure,\footnote{Only one service provider own the newtork in the case of sharing agreements with MVNOs.} without the intervention of a third party \citet{gsma2012}. Examples include roaming agreements between two mobile network operators, sharing of masts or towers between mobile network operators in remote areas, and wholesale broadband network access between an Internet service providers and a vertically integrated fixed broadband network operator.

Network infrastructure is a strategic asset for competition in the downstream market. It determines both the cost and quality of services. As such, service providers owning the network infrastructure may discriminate by raising the access cost or limiting the quality of services of competitors seeking network access under a bilateral agreement. Access regulation is meant to avert such potential discrimination through (i) ex ante interventions such as wholesale access price fixing or mandating equivalence in the quality of inputs or outputs, and (ii) ex post measures pertaining to dispute resolution mechanisms.\footnote{\citet{cave2018disruptive} discussed the role of access regulation in the context of 5G network roll out.} Access regulation has been used as part of cooper local loop unbundling in Europe and several advanced economies with well developed cooper-based telephone networks.

Under partial ownership, telecom service operators typically establish joint ventures among themselves or with a third party specialized in the operations of network infrastructure. For instance, China Tower was formed as a joint venture between the country's three mobile network operators; and MTN, the pan-African mobile network operator, established a joint venture with American Tower Company in Ghana and Uganda in order to share its towers with competitors. Partial ownership can also involves co-investment by rival service providers. Such models were considered as part of the deployment of last-mile fiber optic network in developed economies like France and more generally in Europe. Partial ownership also carries a number of risks, especially coordination failure among service operators with competing interests, potentially resulting in delays in network deployment.

Under a no-ownership scenario, the shared telecom infrastructure is not owned by any service providers.\footnote{This recognizes instances where network operators retain their network infrastructure while entering into sharing agreement on infrastructure own by an independent wholesale access provider.} The typical business model under such scenario involves ownership and operations of the telecom infrastructure by an independent provider - for instance tower companies (hereafter 'towercos') in the case of mobile towers \citep{gvh_etal_2021}. The towerco business model is gaining momentum across developing countries, though there are large disparities across countries and regions. As of 2020, three in four mobile towers in emerging markets were managed by towercos \citep{gvh_etal_2021}. The South East Asia region had the highest share of towers managed by towercos (91 percent), primarily driven by the 100 percent rate in China. This was followed by South Asia (76 percent), primarily driven by 84 percent in India; and Latin America (59 percent), primarily driven by Brazil (70 percent) and Mexico (90 percent).

According to \citet{amadasun2020} the sharing potential in EMs is bigger than in developed economies and more than 87\% of the total CAPEX costs could be shared, with the biggest savings potential in site acquisition and design (41\%), power (31\%) and BTS/NodeB (15\%). In OPEX, 69\% of the totals costs in EMs could be shared, with the biggest potential in software support (20\%), power (20\%), land rent (15\%) and backhaul (14\%).

\section{Related Literature}
\label{sect:literature}

The assessment of the economic benefits of shared telecom infrastructure relates to the broader literature on the welfare effects of telecom market structure. A number of studies have investigated the welfare effects of competition among vertically integrated mobile network operators, i.e., owing the network infrastructure and providing Internet services to end-users. Examples include \citet{houngbonon2016, jeanjean2017, genakos2018} and \citet{elliott2021} who typically find positive impact of competition on price, but negative impact on quality or investment, beyond a certain a level of competition intensity. While integrated MNOs may enter into bilateral infrastructure sharing agreements, these studies did not focus on such arrangements. Other studies like \citet{kim2011} considered the effects of shared mobile infrastructure as part of the regulation of network access by mobile virtual network operators, and generally found that incumbent's investment is not affected under voluntary access regulation, but drop under mandated access regulation.

Several studies have investigated the impact of shared fixed broadband infrastructure (cooper or fiber optic networks) on investment incentives and competition in advanced economies, especially in the EU. Examples include \citet{nardotto2015} who assessed the effect of local loop unbundling (LLU) in the United Kingdom and find that LLU only has a limited positive effect on broadband penetration and the effect disappears after the first years when the market reaches maturity; with a positive impact on service quality via competition. \citet{bourreau2018} investigated the impacts of traditional one-way access obligations and co-investments on the roll/out of network infrastructure and found that access obligations lead to a smaller roll-out of infrastructure due to reduced returns to investment and uneven distribution of investment risks (i.e. the investor bearing all risks compared to the new entrant).

Evidence on the benefits of shared mobile telecom infrastructure remains limited at this stage, especially in developing countries. A recent study by \cite{koutroumpis2021} investigated the impact of shared mobile infrastructure but focused on EU countries and pertain to bilateral sharing. That study found that bilateral infrastructure sharing had resulted in lower prices and improved network coverage and quality for consumers driven by cost reductions, higher returns on investment and increased competition intensity.

Few studies have considered the impact of shared telecom infrastructure in developing countries but most are limited in scope or relied on qualitative approaches or executive surveys. \citet{arakpogun2020} investigated barriers to infrastructure sharing in seven SSA countries based on interviews with different stakeholders and found that incumbents might be reluctant to share networks as this would erode their competitive advantage, and many markets have been dominated by one or two MNOs for years, which lack institutional incentives to engage in infrastructure sharing. \citet{mamushiane2018} quantified the economic impact of infrastructure sharing using Software Defined Networking (SDN) and estimated that in the case of South Africa, full implementation of sharing can reduce the time to recover CAPEX investments and reach profitability from 5.4 months to less than 1.3 months in rural areas, suggesting profitable operation even less economically attractive zones.

Other studies have focused on the benefits of shared infrastructure for mobile network operators. \citet{kim2018} assessed the impact of infrastructure sharing deals on firm performance using a DiD methodology and find statistically significant reductions OPEX for national roaming by 14\% in the short run and by 19\% in the long run. \citet{amadasun2020} estimated that sharing of sites and antennas could reduce CAPEX costs by 20-30\%. Sharing of also the radio network can reduce CAPEX by 25-45\%. Sharing of all the assets could reduce CAPEX by another 10\%.

\section{Conceptual framework}
\label{sect:concept}

Understanding the economic benefits of shared telecom infrastructure through independent towercos would start with a framework to rationalize why mobile network operators divest their towers to an independent operator. The minimization of transaction costs as discussed by \citet{williamson1979} could shed some light on the rationale for such strategy. However, in this paper, we are taking the cost savings for MNOs from shared infrastructure through towercos as given and focus on how such savings can affect competition in the downstream market for retail mobile connectivity, and ultimately the welfare of end-users.

As reported by the literature, shared telecom infrastructure can generate cost savings for MNOs, ranging from 20 to 30 percent \citep{kim2018,amadasun2020}. These wholesale cost savings in the upstream of the mobile connectivity value chain can benefit end-users by (i) increasing the intensity of competition in downstream markets, and (ii) being passed through to end-users in terms of reduced price, depending on the intensity of competition.

Indeed, shared mobile towers dramatically lowers the \textbf{cost of entry} into mobile markets by removing that part of capital expenditure dedicated to network deployment, which is replaced by an operating expenditures in the form of a lease rate to access towers. Under such passive infrastructure sharing scheme, MNOs still incur the capex associated with radio spectrum and base stations - i.e., equipment that connect end-users and manage traffic - but the overall capex is nonetheless reduced by up to 40 percent according to \citet{amadasun2020}. Such reduction in entry cost contributes to level the playing field between large and smaller MNOs, including new entrants, and, therefore, can enable faster network coverage and increased competition intensity.

Shared mobile towers can also reduce the marginal cost of mobile connectivity services by supporting a drop in operating expenditure for MNOs. In particular, site rental cost and energy cost can be shared under the towerco business model, resulting in a drop in operating expenditure. In addition, MNOs also save on maintenance cost of mobile towers as the towerco business model reduces transportation cost to multiple sites. Part of these savings on marginal cost can be passed on to end-users as a drop in price, for a given level of usage/quality, depending on the prevailing intensity of competition in the downstream market.


\section{Data and descriptive statistics}
\label{sect:data}

\subsection{Data and variables}

Data used in this study cover 137 developing countries between 2008 and 2020.\footnote{Developing countries are defined as low and middle-income countries according to the WBG's classification of 2020} We assembled data from five main sources: telecom tower data from TowerXchange (TXC), a leading industry research firm in the tower sector;\footnote{TX is a research institute dedicated to the global telecom tower industry. As of 2022, TowerXchange is a division of Euromoney Global Limited, a publisher of consumer and business journals and periodicals, and is governed with the support and advice of an informal network of advisors composed of executives from the tower and mobile industries.} mobile connectivity data from GSMA, the global association of mobile operators, the International Telecommunications Union (ITU), and the Gallup Survey; and socio-economic data from the World Bank's World Development Indicators database. Table \ref{tab:data} provides the list of variables retrieved from these data sources. Table \ref{tab:sumstat} provides the summary statistics of the main variables.

In order to test the five hypotheses above, we consider the following outcome variables:
\begin{itemize}
    \item Availability of mobile connectivity, measured by the percentage of population covered by at least 3G or 4G mobile network technology. The population coverage data comes from GSMA Intelligence.
    \item Competition intensity, proxied by change in market concentration, which is measured by the Herfindahl Hirschman Index (HHI). We obtained the HHI of both mobile telephony and mobile broadband Internet using market share data from GSMA Intelligence.
    \item Price of mobile connectivity, measured by (i) the price of mobile telephony; and (ii) price of mobile broadband Internet. Price data comes from the ITU's ICT price baskets. More specifically, the price of mobile telephony corresponds to that of the least expensive offer with 70 minutes of voice calls and 20 SMS. The price of mobile broadband Internet represents that of the least expensive offer with 2 GB of data allowance.\footnote{We did not use low and high-usage mobile broadband baskets due to limited historical data -- they only started from 2018.}
    \item Uptake of mobile connectivity is measured by the number of unique mobile telephony and broadband connections (in thousands) and subscribers as a percentage of the population. This data is obtained from GSMA Intelligence, based on the number of subscriptions reported by mobile operators. The number of subscription was adjusted by the number of SIM cards per user to obtain an estimate of the number of unique users.
    \item Digital inclusion is measured along the access dimension, i.e., number of individuals having subscribed to a mobile network. We considered two attributes of inclusion, namely gender and area of residence. In particular, we use the share of women or rural residents with a mobile broadband subscription as our measures of digital inclusion. The Gallup survey focused on access at the household level between 2010 and 2015, before switching to individual-level access from 2016. Our analysis will focus on the 2010-2015 period where most tower transactions took place. 
\end{itemize}

Our treatment variable is a dummy equals to 1 when a tower transaction occurs in a country at a given year, and 0 otherwise. As such, a country can be treated several times due to multiple tower transactions. A tower transaction is defined as the transfer of towers from mobile operators to an independent company. The towerco typically rent access to the acquired towers backed to all mobile operators. Absent a tower transaction, i.e., when the treatment dummy takes the value 0, mobile operators may engage in bilateral tower sharing. However, bilateral sharing of towers is not prevalent in developing countries which is the focus of this paper.

The treatment variable derives from two variables built from TXC:

\begin{itemize}
  \item{\textbf{Annual tower deal size by country.} We built this variable using tower transactions data. TXC publishes annual reports on trends in the tower industry by country, including a summary of mergers and acquisitions, with the number of tower sites involved, the value of the transaction and the type of deals.\footnote{Types of deals include sales and leased back, joint venture, manage with license to lease, portfolio acquisition or transfer} From these reports, we retrieved data on 156 towers acquisition deals between 2008 and 2020 which occurred in 36 developing countries. Tables \ref{tab:deals} and \ref{tab:sites} present the number of deals and the corresponding number of sites by country and year. The median deal over the period and sample countries involved 916 towers (See distribution of deal size in Table \ref{tab:dealsize}).
  
  This data has been collapsed at the country and year level, with a variable $\textit{deal\_sites}$ corresponding to the total number of tower sites involved in deals - that variable equals 0 if no deal was recorded in a given year. It is used at the estimation stage to define a treatment variable based on a minimum number of sites involved in tower deals. The number of deals over the period of 13 years varies by country. Most countries (22/36) had 1 or 2 deals over that period; the remaining 14 countries had 3 to 29 deals (See Table \ref{tab:dealscountry}). This distribution reflects countries with multiple deals within a given year. When these multiple deals are considered as a single 'treatment', the number of treatments over the period range from 1 to 10 by country (See Table \ref{tab:towertreatment}). Our estimation strategy will take into consideration such multiple treatments setup.}
  
  \item{\textbf{Tower sites managed by mobile network operators and towercos.} From the TXC annual report, we retrieved the number of towers managed by mobile operators and towercos across 66 developing countries from 2015 and 2020. The share of towers managed by towercos in a country can drop over the year if the market is adding more towers than they manage. Our dataset comes with missing values, making it difficult to compare the share of towers managed by towercos across years. However, the share fluctuates between 30 and 38 percent (Table \ref{tab:towers}). Some countries like Colombia, India and South Africa experienced significant rise in the share of towers managed by towercos. Among countries with tower data, the average deal size represents 12 percent of the stock of towers (Table \ref{tab:dealshare}).
  }
  
\end{itemize}

The analysis controls for income and market size using socio-economic data on population and GDP assembled from the World Development Indicators database of the World Bank.

\begin{table}[!htbp]\centering
\caption{Summary Statistics}
\label{tab:summary_stats}
\begin{threeparttable}
   \scalebox{.7}{
\begin{tabular}{@{}lrrrrrrrr@{}}\toprule\toprule
Variable & Obs. & Countries & Min Year & Max Year & Mean & SD & Min & Max \\ 
\midrule
\multicolumn{9}{@{}l}{\textit{Panel A: Outcome Variables}}\\
Mobile-voice price, PPP                   & 1\,582 & 159 & 2008 & 2020 & 2.94 & 0.63 & 0.60 & 4.57 \\
Mobile-data price, PPP                      & 1\,043 & 156 & 2013 & 2020 & 3.07 & 0.65 & 0.26 & 5.66 \\
Mobile connections (in 1000)             & 2\,359 & 182 & 2008 & 2020 & 1\,606.24 & 2\,871.19 & 0.03 & 11\,243.64 \\
Mobile subscriptions (population share)                & 2\,359 & 182 & 2008 & 2020 & 0.55 & 0.21 & 0.00 & 0.90 \\
Internet connections (in 1000)          & 2\,062 & 180 & 2008 & 2020 & 606.45 & 1\,186.32 & 0.00 & 4\,625.74 \\
Internet subscriptions (population share)              & 1\,985 & 182 & 2010 & 2020 & 0.28 & 0.20 & 0.00 & 0.82 \\
Internet access, rural households (\%)                   & 628 & 125 & 2010 & 2015 & 0.23 & 0.23 & 0.00 & 1.00 \\
Internet access, women-headed  households (\%)            & 639 & 126 & 2010 & 2015 & 0.29 & 0.25 & 0.00 & 0.92 \\
Retail market HHI (log)                          & 2\,116 & 171 & 2008 & 2020 & 8.34 & 0.31 & 7.21 & 9.20 \\
\addlinespace
\multicolumn{9}{@{}l}{\textit{Panel B: Deal Variables}}\\
Tower-sharing dummy                              &   572 &  44 & 2008 & 2020 & 0.19 & 0.39 & 0.00 & 1.00 \\
Sites transferred in deal                        & 2\,366 & 182 & 2008 & 2020 & 212.17 & 3\,190.89 & 0.00 & 95\,477.00 \\
Total towers in country                          &   335 &  72 & 2015 & 2020 & 57\,125.45 & 251\,374.39 & 749.00 & 2\,094\,464.00 \\
Towers owned by towercos                         &   358 &  72 & 2015 & 2020 & 41\,505.72 & 241\,163.81 & 0.00 & 2\,094\,464.00 \\
\addlinespace
\multicolumn{9}{@{}l}{\textit{Panel C: Control Variables}}\\
Regulatory regime G1                     & 2\,366 & 182 & 2008 & 2020 & 0.19 & 0.39 & 0 & 1 \\
Regulatory regime G2                      & 2\,366 & 182 & 2008 & 2020 & 0.37 & 0.48 & 0 & 1 \\
Regulatory regime G3                      & 2\,366 & 182 & 2008 & 2020 & 0.21 & 0.41 & 0 & 1 \\
Regulatory regime G4                     & 2\,366 & 182 & 2008 & 2020 & 0.09 & 0.29 & 0 & 1 \\
GDP (current US\$)                               & 2\,200 & 175 & 2008 & 2020 & 1.36e11 & 8.46e11 & 2.71e7 & 1.61e13 \\
Population (millions)                            & 2\,361 & 182 & 2008 & 2020 & 34.01 & 144.44 & 0.01 & 1\,441.77 \\
\bottomrule\bottomrule
\end{tabular}
}
\caption*{\footnotesize{\textit{Note: The G1–G4 dummies indicate four mutually exclusive regulatory regimes governing mobile infrastructure. See Table \ref{tab:data} in the Appendix for a glossary of the variables.}}}
\end{threeparttable}
\end{table}

\subsection{Descriptive statistics}



\begin{figure}[!htbp]
  \centering
    \caption{Mean evolution of prices and network outcomes, treated vs.\ never-treated countries}
  \begin{subfigure}[b]{0.48\textwidth}
    \includegraphics[width=\linewidth]{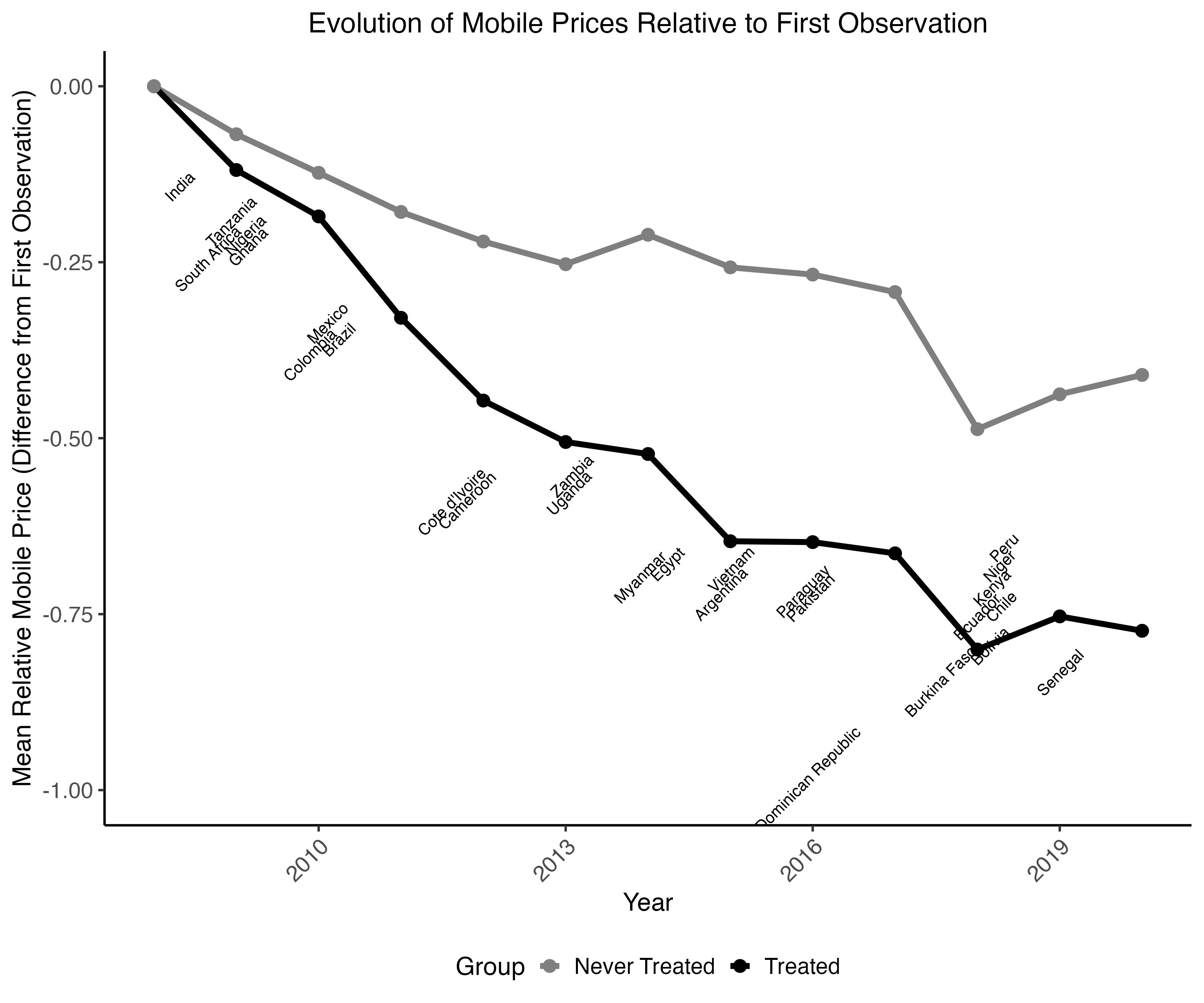}
    \caption{Mobile‐Voice Price (PPP / minute)}\label{fig:rel_p_cell}
  \end{subfigure}
  \hfill
  \begin{subfigure}[b]{0.48\textwidth}
    \includegraphics[width=\linewidth]{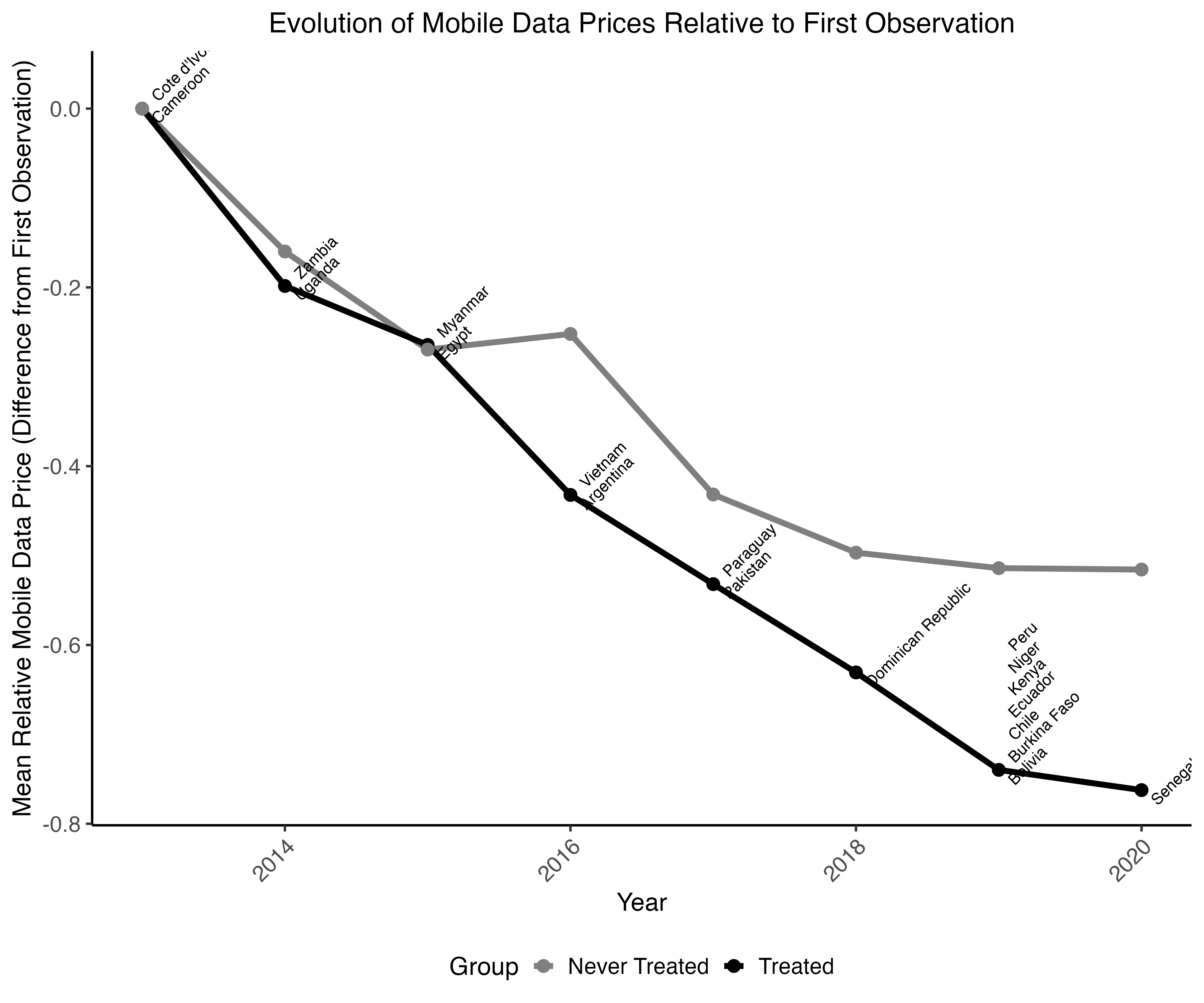}
    \caption{Mobile‐Data Price (PPP / GB)}\label{fig:rel_p_data}
  \end{subfigure}

  \vspace{0.8em}  

  \begin{subfigure}[b]{0.48\textwidth}
    \includegraphics[width=\linewidth]{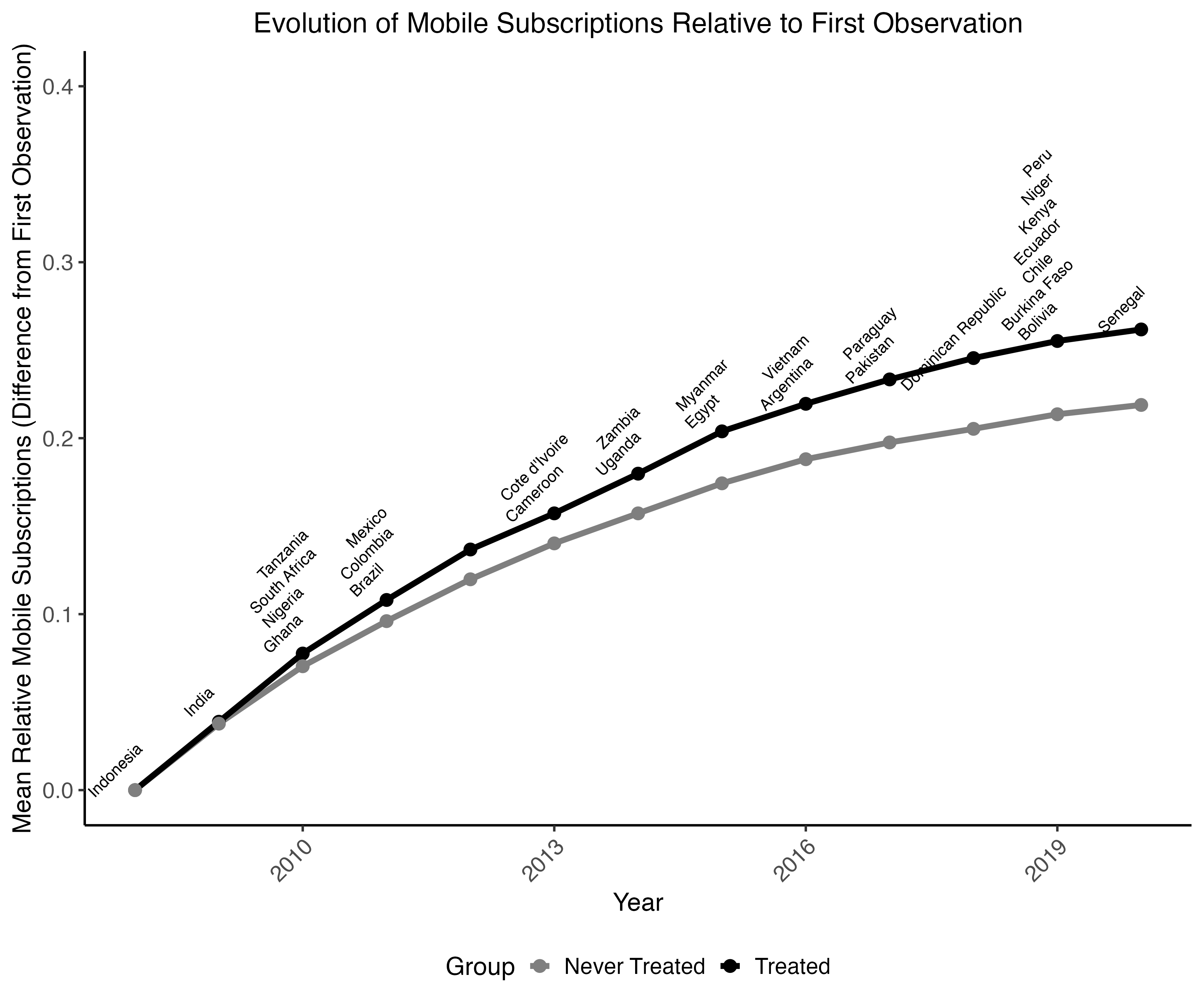}
    \caption{Internet Subscriptions}\label{fig:rel_cov3g}
  \end{subfigure}
  \hfill
  \begin{subfigure}[b]{0.48\textwidth}
    \includegraphics[width=\linewidth]{plots/relative_subscriptions_mobile_treated_untreated.png}
    \caption{Mobile Subscriptions}\label{fig:rel_conn}
  \end{subfigure}

  \captionsetup{justification=raggedright}
\caption*{\footnotesize{\textit{Note: Each panel shows annual cross-country means of the indicated variable.  Series are normalised to zero in the first year each country
    appears in our sample, so the vertical axis measures the \emph{change relative
    to that baseline}.  \textcolor{black}{\textbf{Black}} lines trace countries
    that ever experience a tower-sharing transaction (“treated”); 
    \textcolor{gray}{\textbf{grey}} lines trace those that never do.  Country names
    at top of each panel indicate the year a country was treated.}}}
  \label{fig:rel_2x2}
\end{figure}

Figure \ref{fig:rel_2x2} compares the mean trajectory of prices and network outcomes for the set of countries that ever engage in a tower-sharing transaction (“treated”) and the set that never do (“never-treated”). In every panel the series have been country-demeaned so that the first year of observation equals zero; the vertical axis therefore records the cumulative change with respect to the country-specific baseline. The top row displays price variables—mobile-voice tariffs expressed in PPP cents per minute (Panel (a)) and mobile-data tariffs in PPP dollars per gigabyte (Panel (b)). The bottom row turns to non-price outcomes—3G population coverage (Panel (c)) and a composite index of household internet connectivity (Panel (d)). Black lines trace treated means, grey lines trace never-treated means, and tick-mark labels at the initial point enumerate the constituent countries in order of sample entry. Names of countries at the top of each graph indicate which countries were treated in which year.

The price panels reveal a clear divergence over the sample period. Although prices fall in both groups — a pattern well documented in global mobile markets—the decline is markedly sharper in treated countries. Five years after the start of the time period, the average mobile-voice price in the treated group is about fifteen PPP cents below its baseline, compared with a reduction of roughly seven cents in the never-treated group. The mobile-data gap is even larger: by the end of the sample, treated countries record an average drop of around thirteen PPP dollars per gigabyte, more than twice the decline observed among controls. Taken at face value, these patterns align with the cost-saving mechanism proposed in the tower-sharing literature, whereby lower marginal costs for operators are passed through to consumers as larger price reductions.

Turning to the number of users, both outcomes improve substantially—reflecting the worldwide acceleration in adoption of internet and mobile services—yet the treated series again outpaces the benchmark. Countries that adopted tower-sharing agreements display a larger increase in the number of internet subscribers. The divergence is more pronounced for mobile subscriptions

These figures are descriptive and do not, by themselves, establish a causal impact of tower sharing. They do, however, provide prima facie evidence that our subsequent econometric analysis addresses, as differences in both price and quality measures widen steadily throughout the period.

\section{Econometric Estimation}\label{sect:estimation-results}

\subsection{Difference-in-differences with staggered treatment timing}

Our goal is to quantify the causal impact of a tower-sharing agreement on three outcome families:  
(i) retail prices for mobile voice and data,  
(ii) uptake, measured by mobile- and Internet-connections, and  
(iii) Internet access in traditionally underserved groups—rural and women-headed households. 

Because roll-outs and tariff adjustments unfold gradually, we trace effects in the deal year ($\ell=0$) and in subsequent years ($\ell\ge1$).

\paragraph{Treatment definition.}
Let $\text{Sites}_{it}$ denote the number of tower sites transferred in country $i$ during year $t$.  
We treat a country as \emph{exposed} once it completes a transaction involving at least 1\,000 towers and hold that status thereafter:

\[
G_i \;=\;
\begin{cases}
\displaystyle \min\bigl\{t : \text{Sites}_{it}\ge 1\,000\bigr\}, & 
\text{if such } t \text{ exists},\\[6pt]
\infty, & \text{otherwise},
\end{cases}
\qquad
D_{it} \;=\; \mathbbm{1}\!\{t \ge G_i\}.
\tag{1}
\]
 
Equation~(1) assigns each country a single “first-treatment’’ year, even when follow-up (smaller or larger) transactions arise later.  
Accordingly, the parameters below estimate the average effect of experiencing \emph{at least one} tower deal of 1\,000 sites or more, relative to never having such a deal (or only smaller ones).

\begin{figure}[!htbp]
  \centering
    \caption{Distribution of large tower-sharing deals ($\geq$ 1000 sites)}
  \includegraphics[width=0.75\linewidth]{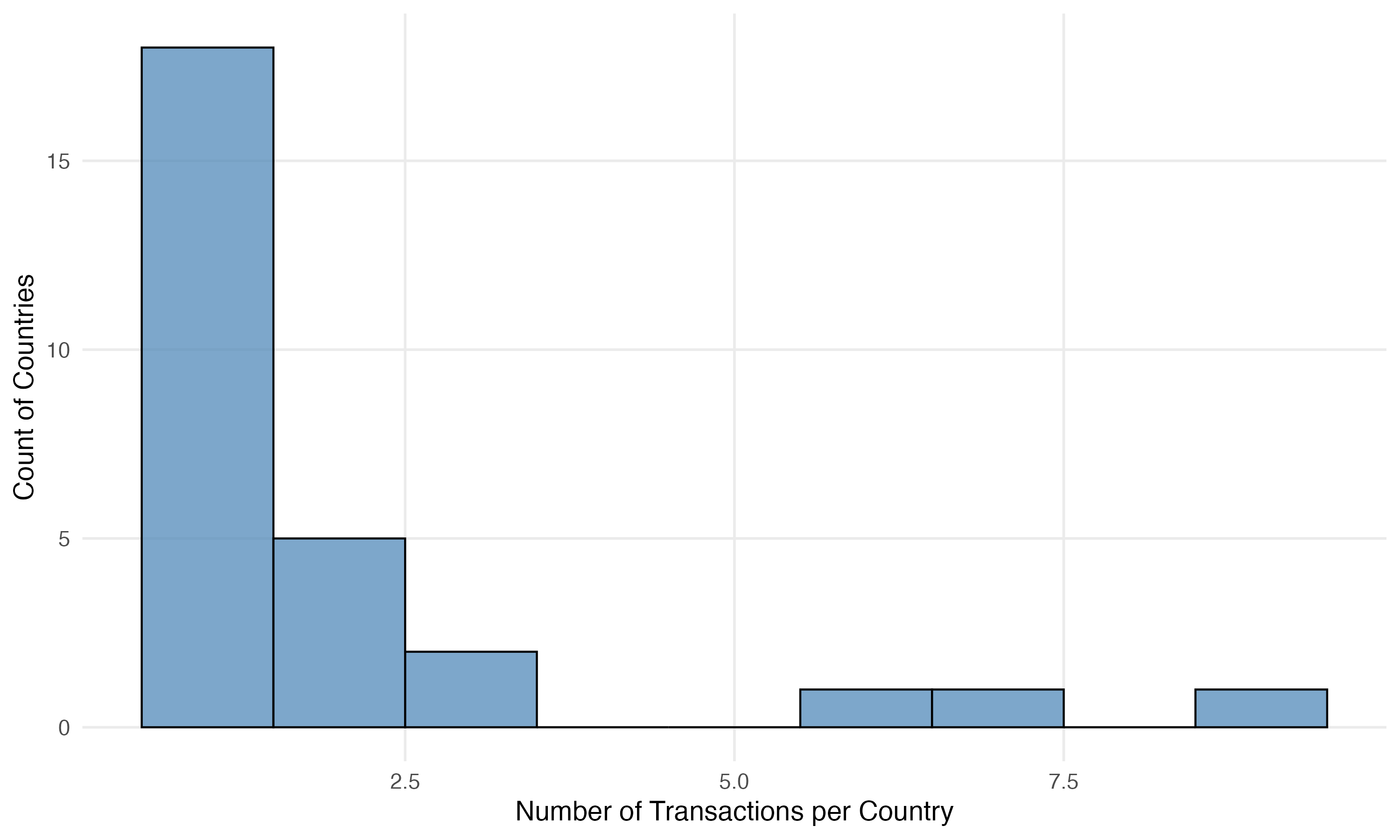}
  \caption*{\footnotesize{\textit{Notes:
 Histogram of the number of qualifying tower-sharing transactions per treated country.
  A “qualifying’’ deal transfers at least 1000 tower sites in a single calendar year.
  }}}
  \label{fig:histogram}
\end{figure}

Under this definition, 28 of the 182 countries in our panel qualify as treated, while the other 154 never reach the 1000-tower threshold. Figure \ref{fig:histogram} plots the number of qualifying transactions per treated country. Most treated countries complete only one large deal; five countries conclude two, and three countries conclude three. The remaining three outliers account for six, seven, and nine qualifying transactions, respectively.

\paragraph{Staggered adoption and dynamic effects.}
Tower-sharing agreements occur in different years across countries, producing a \emph{staggered-treatment} panel.  
We estimate dynamic effects with the interaction-weighted difference-in-differences estimator of \citet*{sun2021estimating}, implemented via \texttt{sunab()} in \textsf{R}.  
Define event time $E_{it}=t-G_i$.  For each relative year $\ell$ we recover the cohort-weighted average treatment effect on the treated,

\[
\widehat{\theta}_{\ell}
\;=\;
\mathbb{E}\!\bigl[Y_{it}(1)-Y_{it}(0)\,\bigm|\,E_{it}=\ell\bigr],
\qquad
\ell\in\{-4,\dots,4\}\setminus\{-1\},
\]

where $\ell=-1$ is the omitted reference year.  
We also report the aggregated post-treatment effect

\[
\widehat{\tau}_{\text{post}}
\;=\;
\frac{1}{L}\sum_{\ell=0}^{L}\widehat{\theta}_{\ell},
\]

with $L$ the maximum observed event time.  Standard errors are clustered by country.

\paragraph{Identification and diagnostics.}
The estimator in \citet*{sun2021estimating} is consistent under  
(i) \emph{parallel pre-treatment trends},  
(ii) \emph{no anticipatory behaviour}, and  
(iii) a \emph{common post-treatment dynamic profile} across cohorts.  
We probe these conditions with placebo (pre-trend) tests and robustness checks reported below.

\paragraph{Sample definition.}
The baseline sample keeps only transactions of at least 1\,000 towers—a scale regarded as the minimum for stand-alone profitability.

\subsection{Results}\label{sect:results}
We begin by examining how tower-sharing agreements affect two first-order
market outcomes: \textit{(i)} retail prices for voice and data and
\textit{(ii)} mobile-connectivity uptake.  
Prices reveal the extent to which cost savings are passed through to consumers,
whereas quantities capture the demand response driven by improved coverage and
lower marginal costs.  
Tables~\ref{tab:prices} and~\ref{tab:uptake} report aggregated post-treatment
effects alongside robustness checks that successively exclude the deal year
and the first two post-treatment years.

Suppressing the contemporaneous and first-post periods serves two purposes.
First, it buffers against anticipatory or transitory shocks in the contract
year—e.g.\ integration costs temporarily loaded onto retail tariffs.
Second, it rejects the possibility that the long-run trajectory merely
reflects mean reversion from an unusually high pre-deal price level.
The persistence of estimated effects across windows therefore reinforces the
interpretation of tower sharing as a medium-run driver of cheaper and more
widely consumed mobile services.

Formally, each coefficient reported in Tables~\ref{tab:prices}
and~\ref{tab:uptake} is
\[
  \widehat{\text{ATE}}
  \;=\;
  \sum_{\ell\ge 0} w_{\ell}\,\widehat{\theta}_{\ell},
\]
where $w_{\ell}$ is the share of treated‐group observations that contribute at
horizon~$\ell$.  
Under the assumptions stated above, these estimates can be interpreted as
causal effects relative to not-yet-treated and never-treated controls.
Standard errors are clustered at the country level throughout.

\paragraph{Prices}
Table~\ref{tab:prices} shows a pronounced decline in mobile‐voice prices following entry.  The baseline ATE implies a -0.58 log points reduction (s.e.\ 0.13, $p<0.01$).  The effect remains economically large and statistically significant when the transaction year is omitted (\SI{-0.43}{}, s.e.\ 0.08) or when the first two post‐treatment years are excluded (\SI{-0.33}{}, s.e.\ 0.10), indicating that short‐lived launch promotions or billing adjustments do not drive it.  For mobile‐data services, the unconditional ATE is imprecisely estimated, but once the early post‐agreement window is removed the price response becomes sizable (-0.36 and-0.53), significant at the 5\% level.  These results point to a sustained pass‐through of cost savings to consumers, with data prices adjusting more gradually than voice tariffs.

\begin{table}[!h]
\centering
\caption{Effects on Mobile Service Prices}\label{tab:prices}
\centering
\begin{tabular}[t]{lccc}
\toprule
\toprule
Outcome & ATE & SE & p-value\\
\midrule
Mobile-Voice Price & -0.578*** & 0.13 & $<$0.001\\
Mobile-Data Price & -0.131 & 0.16 & 0.414\\
Mobile-Voice Price (exclude transaction year) & -0.429*** & 0.08 & $<$0.001\\
Mobile-Data Price (exclude transaction year)  & -0.363** & 0.17 & 0.032\\
Mobile-Voice Price (exclude 2 years)  & -0.329*** & 0.10 & 0.001\\
Mobile-Data Price (exclude 2 years)& -0.526*** & 0.15 & 0.001\\
\bottomrule
\bottomrule
\end{tabular}
\caption*{\footnotesize\textit{Notes}: Each entry reports the average post-treatment effect (ATE) from \cite{sun2021estimating} event-study regression. All specifications include country and year fixed effects and control for GDP per capita, total population and generation fixed effects. “Exclude transaction year’’ drops the event-time coefficient from the post-treatment average; “Exclude 2 years’’ drops the transaction year and the subsequent year. Standard errors, in parentheses, are clustered at the country level.}
\end{table}

\paragraph{Uptake of mobile connectivity}

Table~\ref{tab:uptake} documents pronounced quantity responses. Across all post‐treatment years, the average treated country records an additional \num{142} mobile subscriptions per \num{100.000} inhabitants (s.e.\ 38.6, $p<0.01$).  The effect remains highly significant---and even grows in magnitude---once the transaction year and the first post year are excluded, indicating that user adoption accelerates rather than dissipates after the infrastructure is in place.  For internet connectivity, the baseline estimate is positive but imprecise; once the early window is removed, the effect becomes large (72 and 92) and statistically significant, consistent with households upgrading data plans only after network quality perceptibly improves.

\begin{table}[!h]
\centering
\caption{Effects on Mobile and Internet Connectivity}\label{tab:uptake}
\centering
\begin{tabular}[t]{lccc}
\toprule
\toprule
Outcome & ATE & SE & p-value\\
\midrule
Mobile Connectivity & 1422.412*** & 385.74 & $<$ 0.001\\
Internet Connectivity & 599.743 & 477.18 & 0.210\\
Mobile Connectivity (exclude transcation year) & 934.588*** & 244.30 & $<$ 0.001\\
Internet Connectivity (exclude transcation year) & 916.345*** & 274.34 & 0.001\\
Mobile Connectivity (exclude 2 years) & 643.517*** & 170.69 & $<$0.001\\
Internet Connectivity (exclude 2 years) & 717.252*** & 273.38 & 0.009\\
\bottomrule
\bottomrule
\end{tabular}
\caption*{\footnotesize\textit{Notes}: Each entry reports the average post-treatment effect (ATE) from \cite{sun2021estimating} event-study regression. All specifications include country and year fixed effects and control for GDP per capita, total population and generation fixed effects. “Exclude transaction year’’ drops the event-time coefficient from the post-treatment average; “Exclude 2 years’’ drops the transaction year and the subsequent year. Standard errors, in parentheses, are clustered at the country level.}
\end{table}

\subsection{Dynamic Treatment Effects}

\begin{figure}[!htbp]
  \centering
  \caption{Dynamic treatment effects of tower–sharing entry}

  \begin{subfigure}[b]{0.48\textwidth}
    \includegraphics[width=\linewidth]{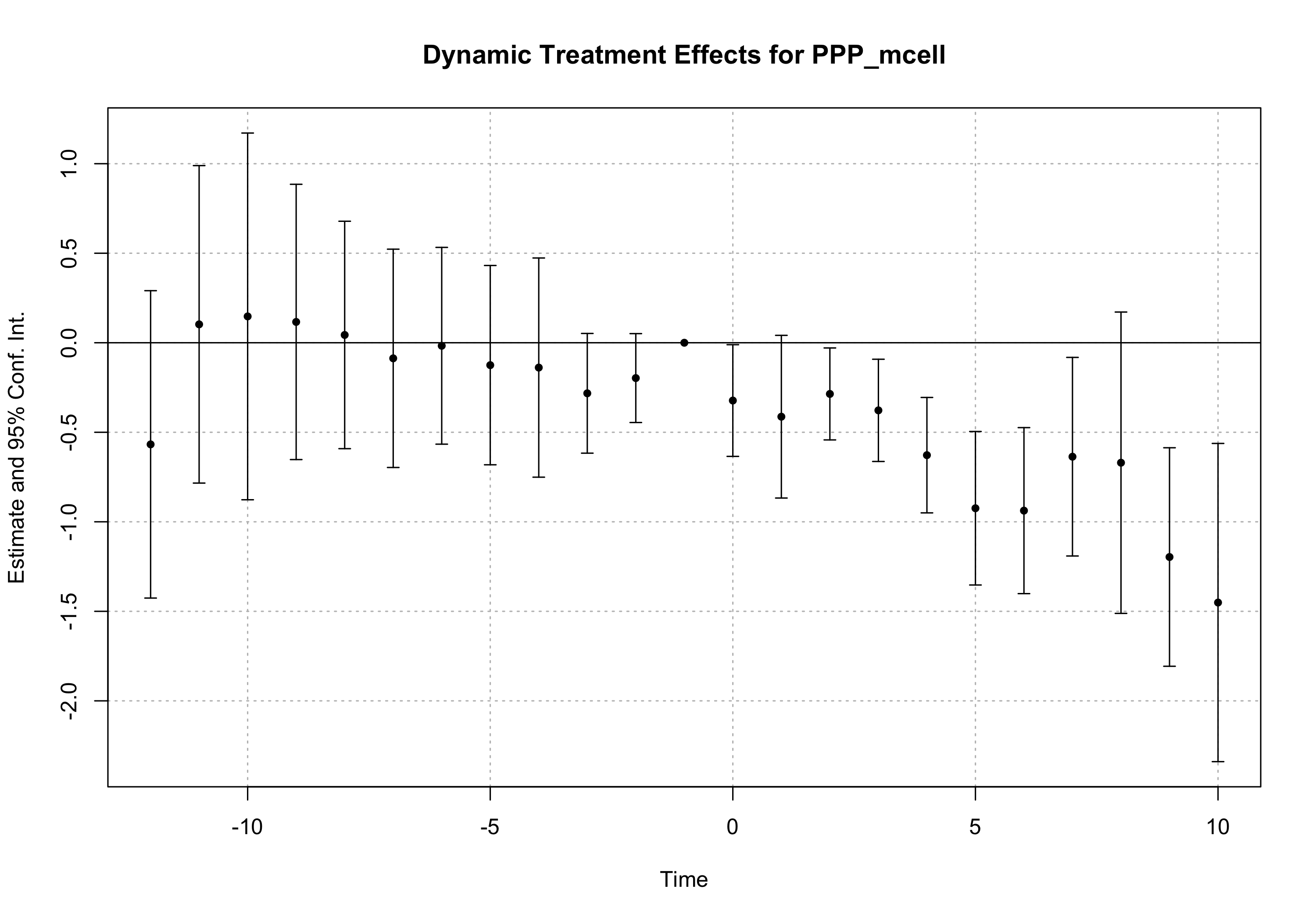}
    \caption{Mobile–voice price (PPP / minute)}\label{fig:dte_ppp_mcell}
  \end{subfigure}
  \hfill
  \begin{subfigure}[b]{0.48\textwidth}
    \includegraphics[width=\linewidth]{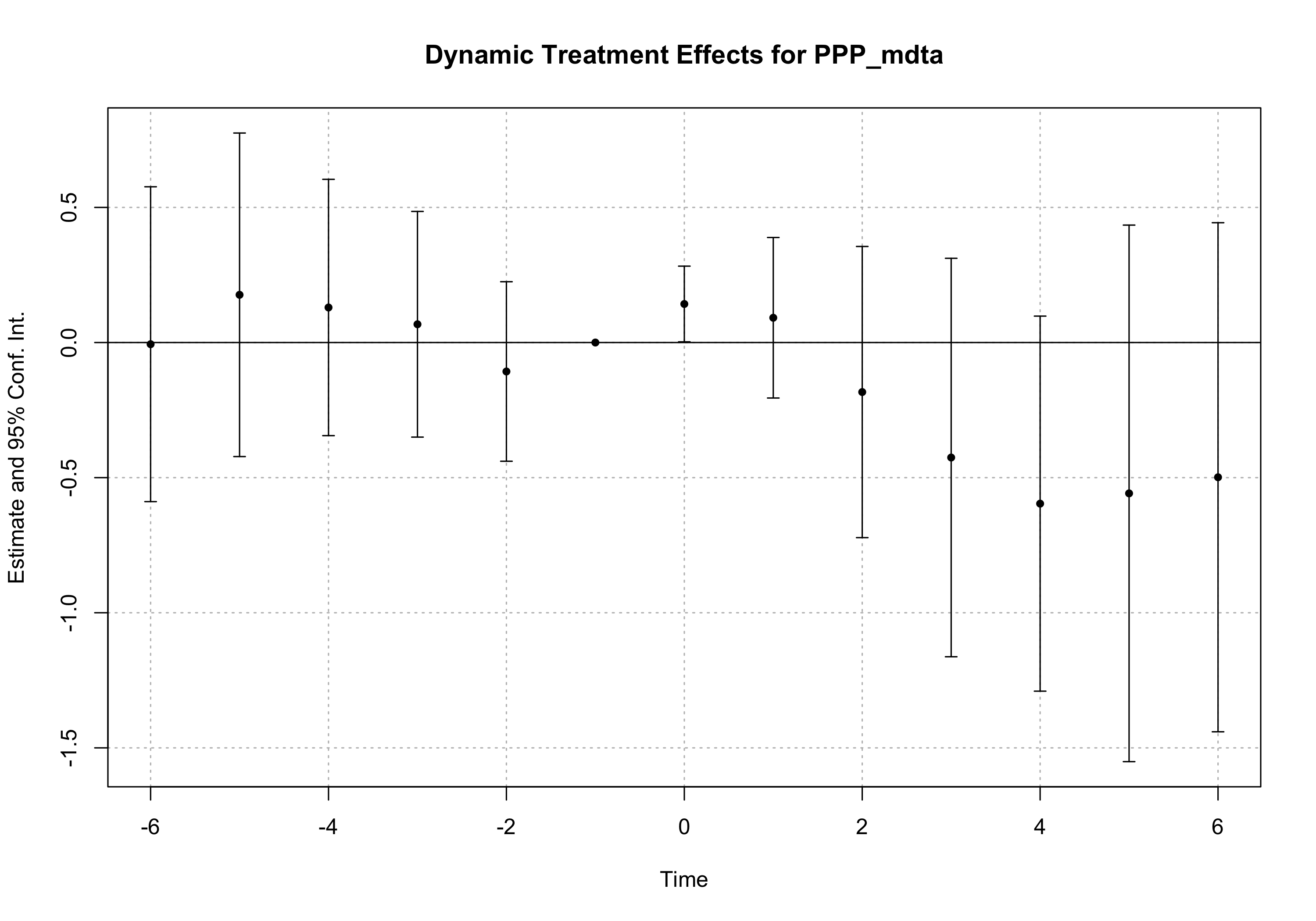}
    \caption{Mobile–data price (PPP / GB)}\label{fig:dte_ppp_mdta}
  \end{subfigure}

  \vspace{0.8em}  

  \begin{subfigure}[b]{0.48\textwidth}
    \includegraphics[width=\linewidth]{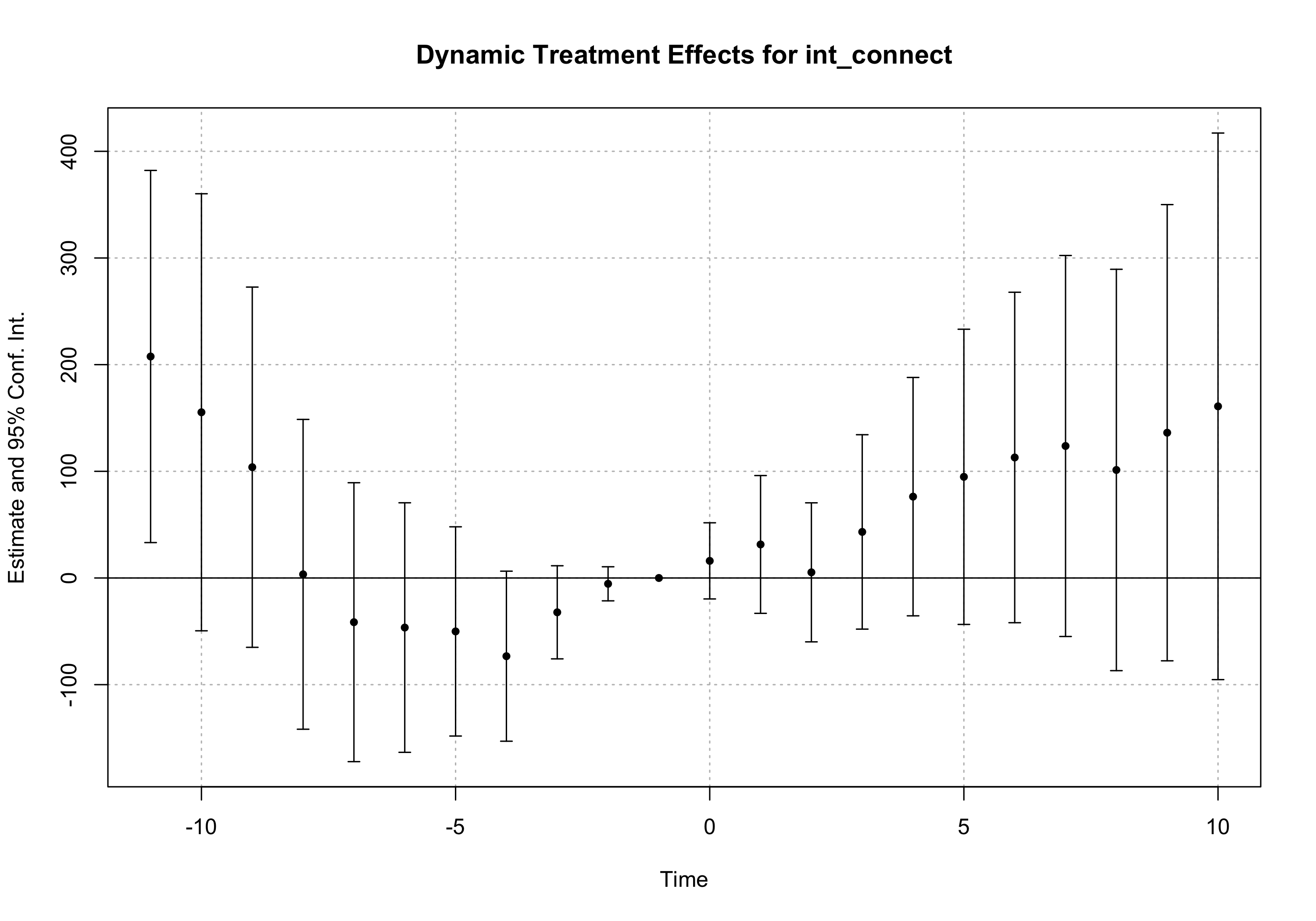}
    \caption{Internet connections }\label{fig:dte_int_connect}
  \end{subfigure}
  \hfill
  \begin{subfigure}[b]{0.48\textwidth}
    \includegraphics[width=\linewidth]{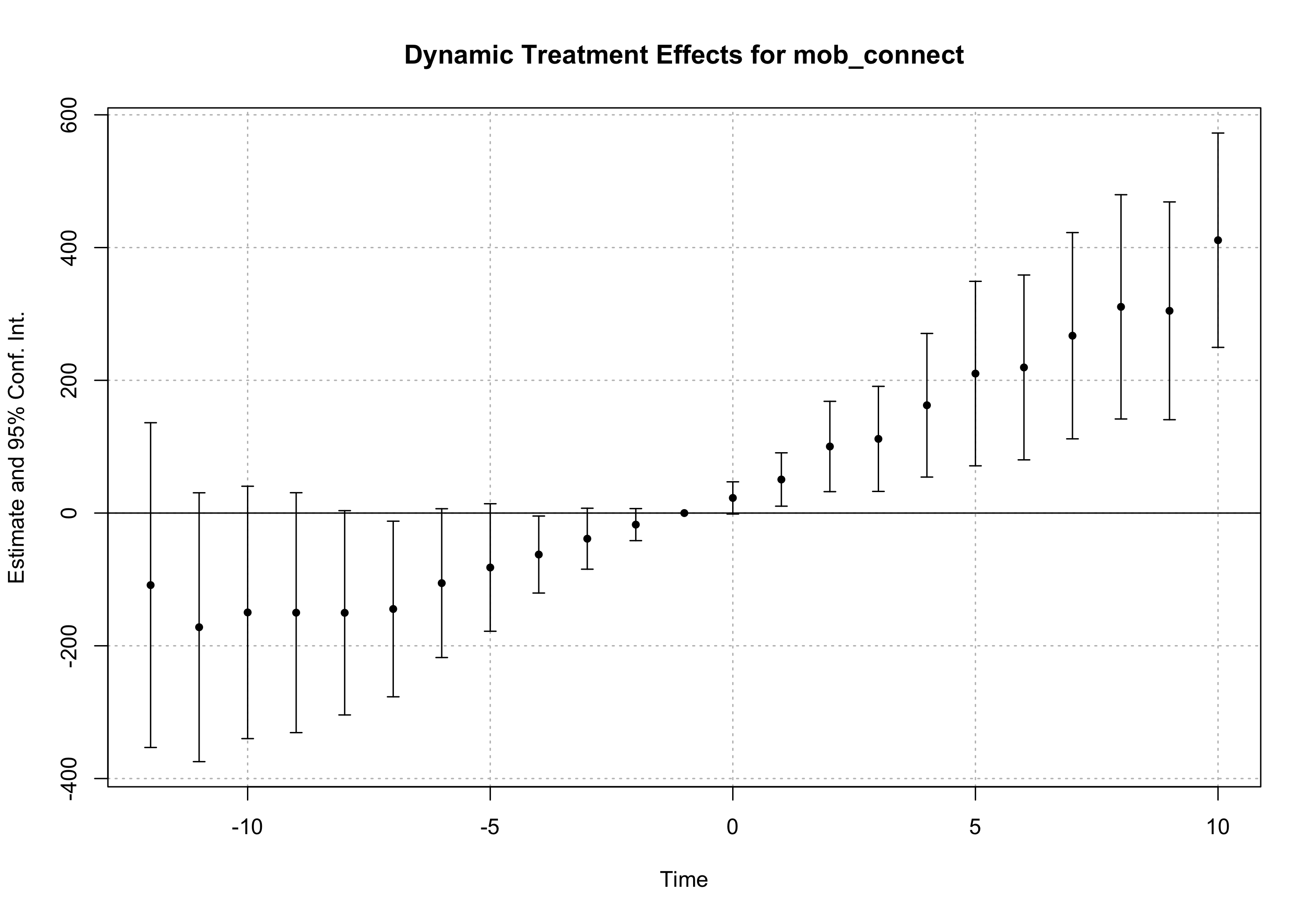}
    \caption{Mobile connections)}\label{fig:dte_mob_connect}
  \end{subfigure}

  \captionsetup{justification=raggedright}
  \caption*{\footnotesize\textit{Notes}: Each panel plots the event‐study
  coefficients $\hat\beta_{\ell}$ obtained with the Sun--Abraham estimator
  (\citeyear{sun2021estimating}).  The horizontal axis is event time
  $\ell$ (years relative to the first tower‐sharing agreement in a
  country), with $\ell=0$ marking the transaction year; negative values
  are leads (placebo periods) and positive values are lags
  (post‐treatment years).  Dots are point estimates and vertical bars
  are 95\% confidence intervals based on country‐clustered standard
  errors.  A flat pre‐period validates the parallel‐trends assumption,
  while post‐period dynamics reveal the magnitude and timing of the
  treatment impact.}
  \label{fig:dte_2x2}
\end{figure}

While the average post–treatment effects in Tables~\ref{tab:prices} and~\ref{tab:uptake} are policy–relevant summaries, they mask the timing of the adjustment.  A central concern with staggered Difference‐in‐Differences designs is that violations of the \emph{parallel‐trends} assumption may occur only in certain periods.  Estimating the event‐ study coefficients $\hat\beta_{\ell}$ therefore serves a dual purpose:(\emph{i})~it provides a placebo test of pre‐treatment equality, and (\emph{ii})~it uncovers whether the treatment effect arrives immediately or accumulates gradually.  Figure~\ref{fig:dte_2x2} plots these dynamics for the four headline outcomes.

\paragraph{Placebo evidence.}
Across all panels, the leads ($\ell<0$) oscillate around zero and are jointly insignificant at conventional levels (Wald $p>0.20$ in every case).  The point estimates are small relative to their post‐treatment counterparts and the 95~percent confidence bands comfortably straddle the horizontal axis.  This absence of systematic pre‐trends lends credence to the identifying assumption that treated and control countries would have continued on parallel paths in the absence of tower sharing.

\paragraph{Timing and magnitude of the treatment response.}
Turning to the lags ($\ell\ge 0$), the price panels display a clear and divergent pattern.  Mobile‐voice tariffs fall almost immediately upon the agreement and continue to decline for at least six years, stabilising at roughly half a PPP dollar below the pre‐treatment baseline.  Mobile‐data prices exhibit a slower ramp‐up: the first two post years are statistically indistinguishable from zero, but sizeable reductions emerge from year~3 onward, eventually converging to a (-0.5 to -0.6) decrease.  The quantity responses mirror this lag structure.  Mobile subscriptions begin to climb within one to two years after the deal, reaching an additional \num{350} subscriptions per \num{100.000} inhabitants by year~10.  Internet subscriptions react even more gradually; significant gains materialise only after three years but continue to accumulate throughout the horizon.

\subsection{Threshold robustness.}
\begin{figure}[!htbp]
  \centering
  \caption{Robustness to alternative tower-count thresholds}

  \begin{subfigure}[b]{0.48\textwidth}
    \includegraphics[width=\linewidth]{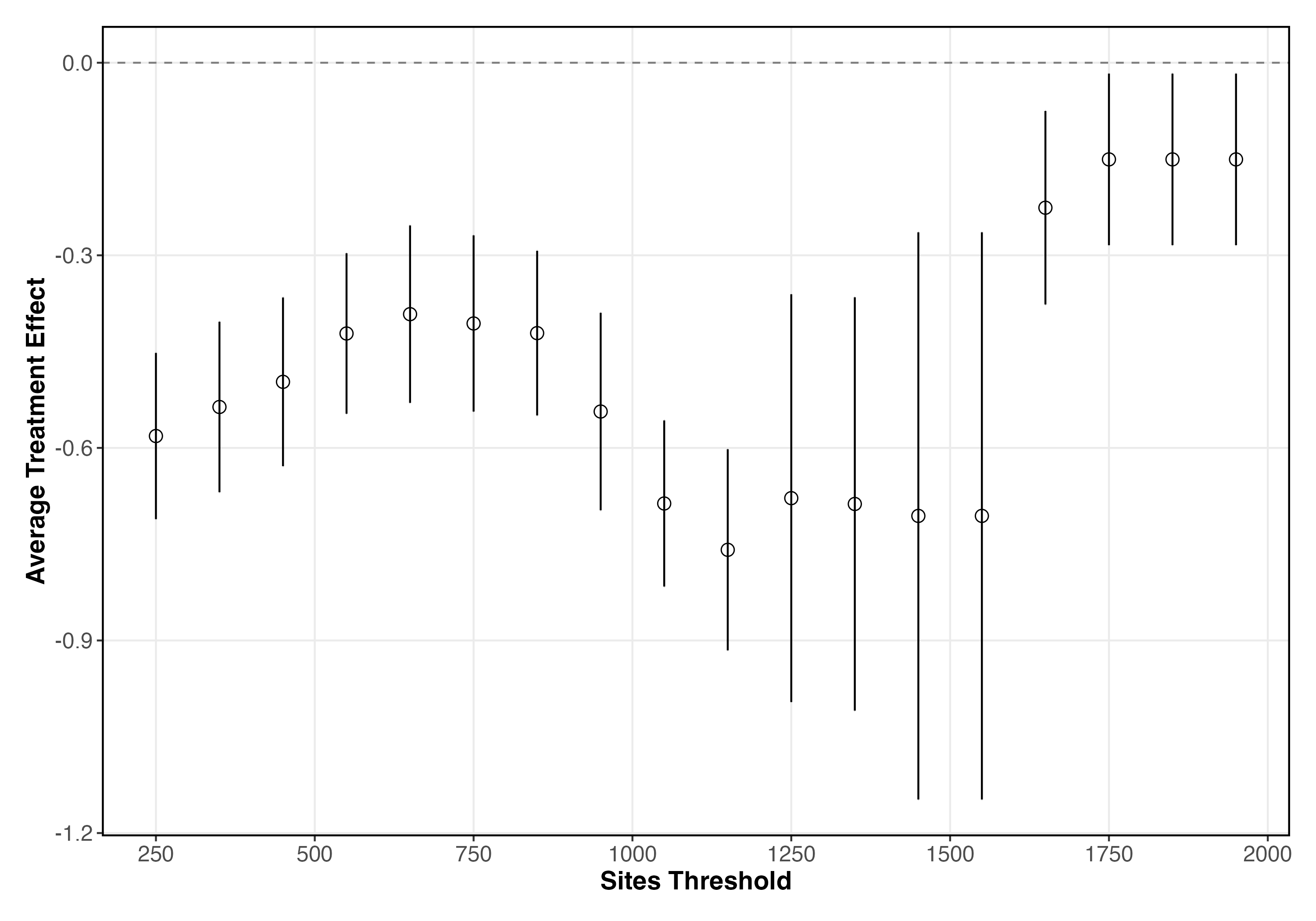}
    \caption{Mobile–voice price (PPP / minute)}\label{fig:thr_ppp_mcell}
  \end{subfigure}
  \hfill
  \begin{subfigure}[b]{0.48\textwidth}
    \includegraphics[width=\linewidth]{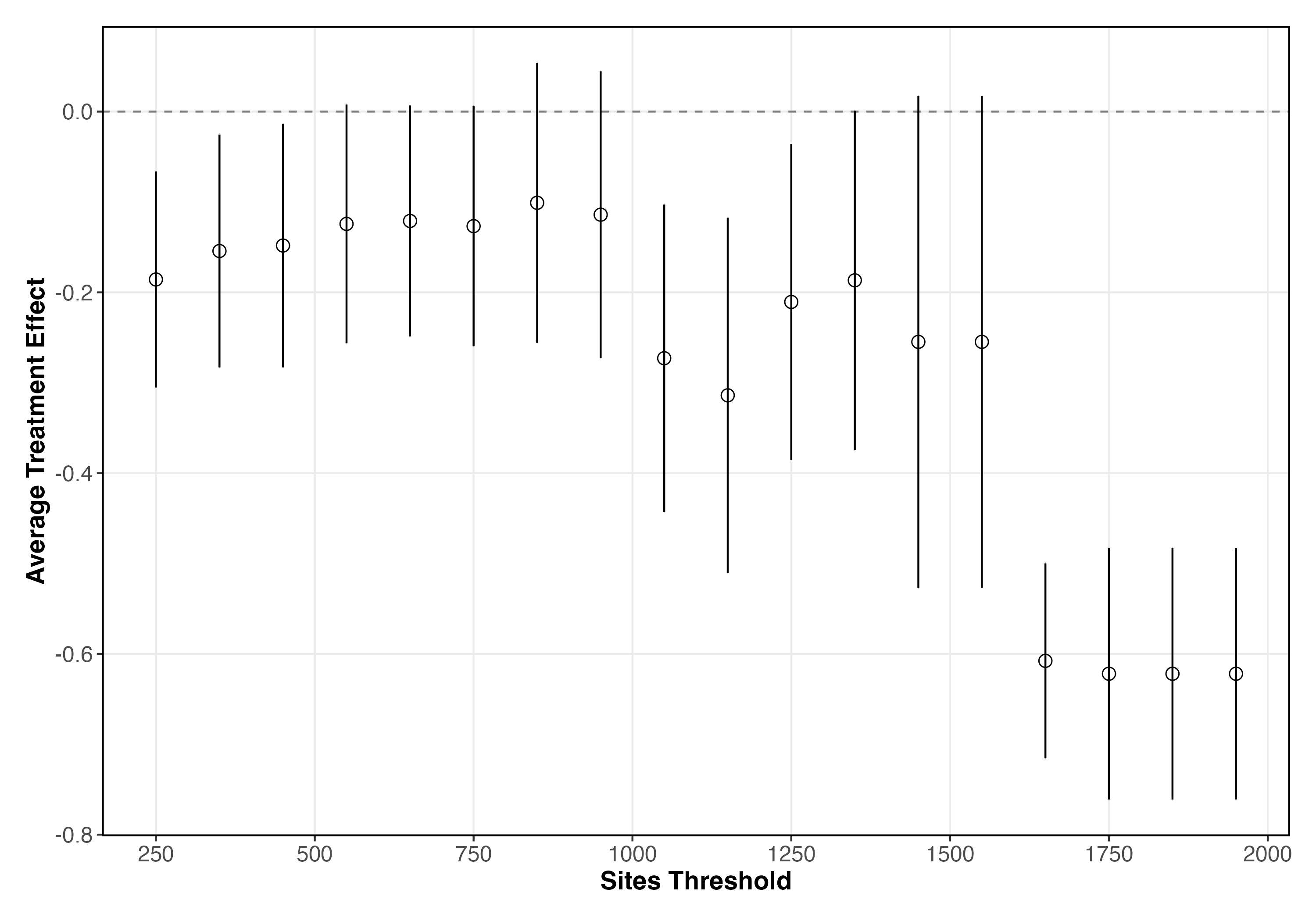}
    \caption{Mobile–data price (PPP / GB)}\label{fig:thr_ppp_mdta}
  \end{subfigure}

  \vspace{0.8em}

  \begin{subfigure}[b]{0.48\textwidth}
    \includegraphics[width=\linewidth]{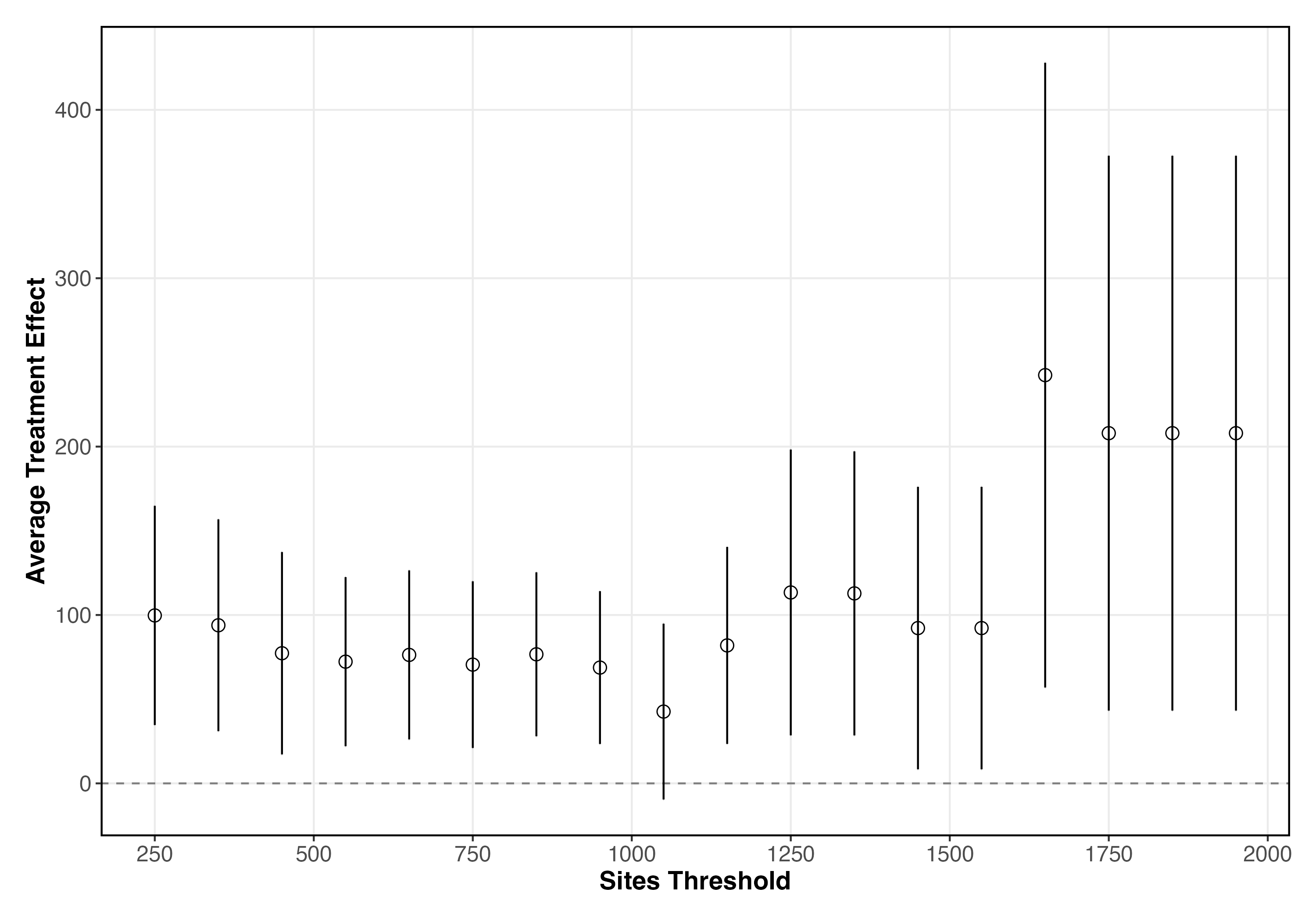}
    \caption{Internet connections}\label{fig:thr_int}
  \end{subfigure}
  \hfill
  \begin{subfigure}[b]{0.48\textwidth}
    \includegraphics[width=\linewidth]{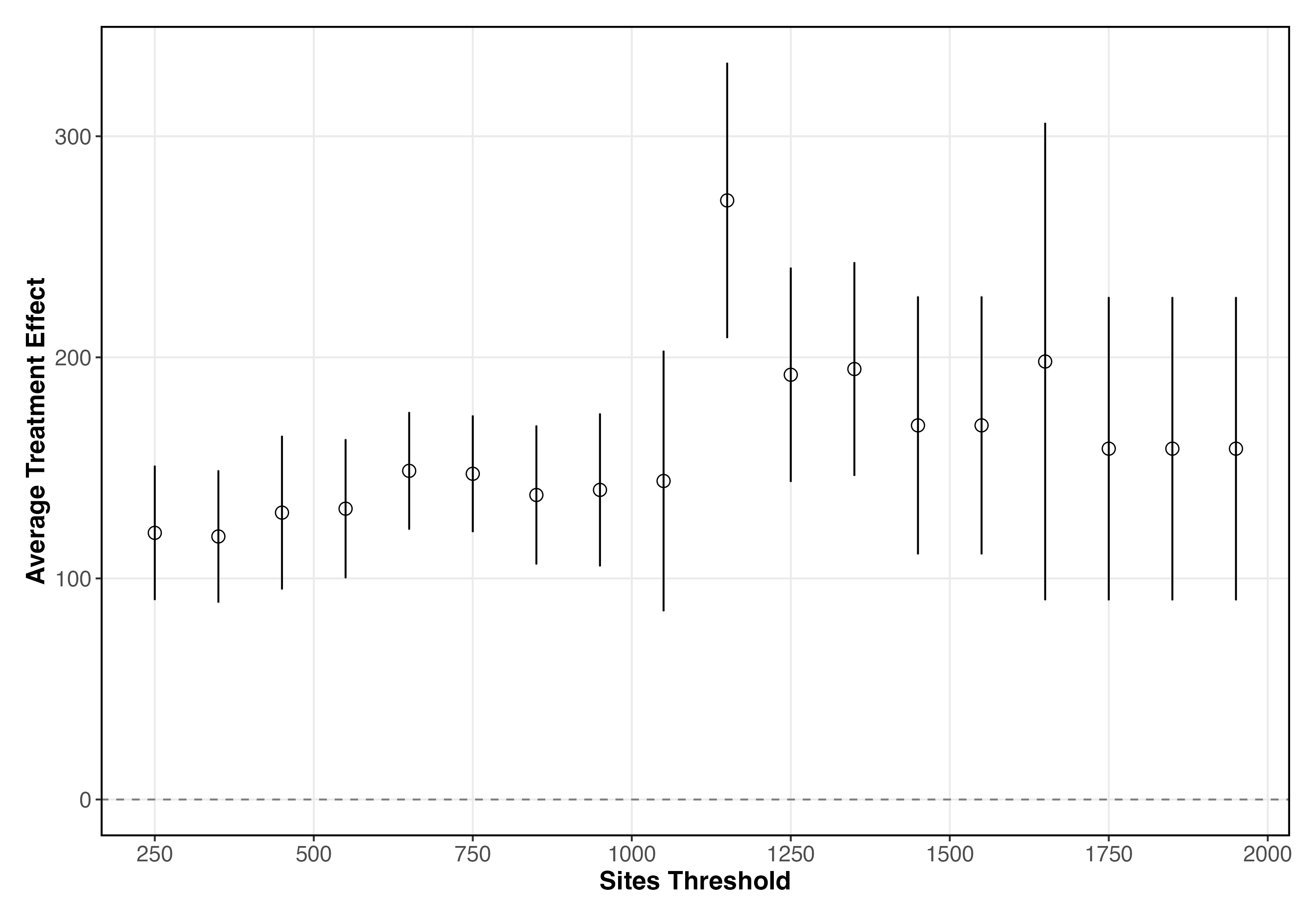}
    \caption{Mobile connections}\label{fig:thr_mob}
  \end{subfigure}

  \captionsetup{justification=raggedright}
  \caption*{\footnotesize\textit{Notes}:  Each marker is the average
  post-treatment effect obtained when the “treated’’ dummy is defined as
  transactions that transfer more than the number of sites shown on the
  horizontal axis.  Vertical bars are 95\% confidence intervals based on
  country-clustered standard errors.  The dashed line marks a zero
  effect; values below (above) imply price reductions (quantity
  increases).}
  \label{fig:threshold_robust}
\end{figure}

Our baseline definition treats an agreement as “tower sharing’’ only if
it transfers at least \num{1000} sites, a threshold that industry
analysts regard as the \emph{minimum viable network}.  Although this
rule is grounded in qualitative expertise, it remains somewhat ad~hoc.
To demonstrate that our main conclusions do not hinge on this single
cut-off, we re-estimate the average post-treatment effect (ATE) for a
grid of alternative thresholds ranging from \num{250} to
\num{2000}~sites.  For each threshold~$\tau$ we recode the treatment
indicator to equal one if, and only if, the transaction involves more
than~$\tau$ sites, and then re-run the Sun–Abraham aggregation.  Figure~\ref{fig:threshold_robust}
plots the resulting ATEs and their 95~percent confidence intervals for
the four headline outcomes.

Two patterns emerge.  First, the sign of the treatment effect is
\emph{stable across thresholds}.  Voice and data prices are always
negative, indicating cheaper services, while subscription measures are
always positive, signalling higher connectivity.  Second, the
\emph{magnitude} of the effect is remarkably flat for thresholds up to
about \num{1000}–\num{1250} sites and then begins to fan out.  Larger
cut-offs generate stronger point estimates but also wider confidence
bands, reflecting the fact that very large transactions are rare and
thus estimated with less precision.  Importantly, even at the top end
of the range the price effects remain economically large
(\SIrange{-0.3}{-0.6}) and the quantity effects exceed
\num{150} additional subscriptions per \num{100.000} inhabitants.

In short, the core conclusions of Tables ~\ref{tab:prices} and ~\ref{tab:uptake} survive a
wide array of alternative tower-count definitions.  The agreement size
of \num{1000} sites used in our baseline lies in a region where the
estimated effects are both precisely measured and representative of the
broader pattern, lending further credibility to the choice of this
threshold.

\subsection{Mechanisms}

\subsection{Exploring Mechanisms}

Why might tower–sharing agreements translate into lower prices and
higher take-up?  We investigate two candidate channels.  First, by
lowering the fixed cost of network expansion, sharing may
disproportionately benefit populations that have historically
lagged in digital adoption—in particular, rural communities and
households headed by women.  Second, sharing can reduce barriers to
entry and thereby intensify market competition.  We probe the
latter mechanism through changes in the Herfindahl–Hirschman Index
(\textsc{HHI}) of retail market shares.

\begin{table}[!htbp]
\centering
\caption{Tower sharing, underserved groups, and market concentration}
\label{tab:mechanisms}
\begin{threeparttable}
\begin{tabular}{@{}lccc@{}}
\toprule
\toprule
\textbf{Outcome} & \textbf{ATE} & \textbf{SE} & \textbf{$p$--value}\\
\midrule
\multicolumn{4}{@{}l}{\textit{Panel A: Internet access in underserved groups (pp)}}\\
\addlinespace
Internet, rural households                         & 0.010 & 0.020 & 0.564\\
Internet, rural households                  & 0.032$^{\ast}$ & 0.020 & 0.057\\
\quad $\;$exclude transaction year                 & 0.047$^{\ast\ast\ast}$ & 0.010 & $<\!0.001$\\
\quad $\;$exclude two years                        & 0.040 & 0.030 & 0.122\\
Internet, women‐headed households                  &       &       &      \\
\quad $\;$exclude transaction year                 & 0.037$^{\ast\ast}$ & 0.010 & 0.015\\
\quad $\;$exclude two years                        & 0.036$^{\ast\ast\ast}$ & 0.010 & 0.003\\
\addlinespace
\multicolumn{4}{@{}l}{\textit{Panel B: Market concentration (Herfindahl–Hirschman Index)}}\\
\addlinespace
HHI                                                & $-0.053^{\ast\ast}$ & 0.020 & 0.024\\
\quad $\;$exclude transaction year                 & $-0.070^{\ast\ast\ast}$ & 0.030 & 0.007\\
\quad $\;$exclude two years                        & $-0.080^{\ast\ast}$ & 0.030 & 0.011\\
\bottomrule
\bottomrule
\end{tabular}
\footnotesize \textit{Notes}: Each entry reports the average post-treatment effect
(ATE) obtained from the Sun–Abraham event-study estimator.  Standard errors in
parentheses are clustered at the country level.  Regressions control for country
and year fixed effects, GDP per capita, total population, and generation
dummies.  ``Exclude transaction year’’ omits event time $\ell=0$ from the
post-treatment average; ``exclude two years’’ omits $\ell=0$ and $\ell=1$.
$^{\ast}$\,$p<0.10$, $^{\ast\ast}$\,$p<0.05$, $^{\ast\ast\ast}$\,$p<0.01$.
\end{threeparttable}
\end{table}

Panel~A of Table~\ref{tab:mechanisms} reports the average
post-treatment effect on internet access among rural and
women-headed households.  Point estimates are positive in all
specifications.  For women-headed households the increase is both
statistically and economically meaningful: a
\SI{3.2}{percentage-point} gain in the baseline, which persists at
\SI{3.6}{pp} when we exclude the transaction year and remains
significant at the 1~percent level after omitting the first two
post-treatment years.  The effect for rural households is concentrated
in the transaction year itself (\SI{4.7}{pp}, $p<0.01$) and fades once
two years have elapsed.

\paragraph{Competition as a conduit.}
Panel~B examines the \textsc{HHI}.  Across all windows tower sharing
lowers market concentration by 5–8 points, with the largest decline
observed when the transaction year is removed
($-0.07$, $p<0.01$).  Because the index ranges from~0 (perfect
competition) to~1 (monopoly), a reduction of this magnitude implies a
non-trivial increase in the number—or at least the effective
symmetry—of active providers.  The timing aligns with the price results:
competitive pressure intensifies quickly after the agreement, helping
explain the swift pass-through to tariffs.

\subsection{Heterogeneous Treatment Effects}\label{sec:hete}

Tower-sharing deals differ in their \emph{contractual form} and \emph{scale}.  
A \textit{sale–and–leaseback} (SLB) transaction, in which the mobile-network operator (MNO) sells its towers to an infrastructure provider and immediately leases back access, leaves the operator as an anchor tenant and preserves coordination incentives.  
By contrast, an outright \textit{acquisition} transfers ownership—and often operational control—entirely to the tower company. These are the two most common types of transactions in our dataset, with SLB accounting for 54\% and acquisitions for 38\%.

Likewise, the impact of the transaction might differ depending on the size of the transaction. To test these hypotheses, we split the treated sample along these two dimensions, estimate group-specific average treatment effects (ATEs), and report the difference in effects across groups.

\begin{table}[!htbp]\centering
\caption{Heterogeneous Treatment Effects by Contract Type and Deal Size}
\label{tab:hete}
\centering
\begin{tabular}{@{}lrrrrrrr@{}}
\toprule\toprule
 & \multicolumn{2}{c}{Group 1} & \multicolumn{2}{c}{Group 2} & \multicolumn{3}{c}{Difference} \\
\cmidrule(lr){2-3}\cmidrule(lr){4-5}\cmidrule(l){6-8}
Outcome & ATE & SE & ATE & SE & $\Delta$ATE & SE & $p$--value \\
\midrule
\multicolumn{8}{@{}l}{\textit{Panel~A: Contract Type — SLB (Group 1) vs.\ Acquisition (Group 2)}}\\
Mobile-voice price       & $-1.065$ & (0.347) &  $0.920$ & (0.497) & $-1.985$ & (0.607) & 0.001 \\
Mobile-data price        &  $0.108$ & (0.301) & $-0.187$ & (0.391) &  $0.296$ & (0.494) & 0.550 \\
Internet connectivity     & $232.657$& (76.178)&  $29.426$& (96.824)& $203.231$&(123.199)& 0.100 \\
Mobile connectivity                & $-130.730$&(133.300)& $-65.743$&(64.399)& $-64.987$&(148.041)& 0.661 \\
\addlinespace
\multicolumn{8}{@{}l}{\textit{Panel~B: Deal Size — $\le$2{,}000 Towers (Group 1) vs.\ $>$2{,}000 Towers (Group 2)}}\\
Mobile-voice price         & $-0.973$ & (0.851) &  $0.634$ & (0.472) & $-1.607$ & (0.973) & 0.100 \\
Mobile-data price         &  $0.025$ & (0.284) & $-0.048$ & (0.299) &  $0.072$ & (0.412) & 0.861 \\
Internet connectivity       & $153.345$& (81.842)& $192.507$& (45.727)& $-39.162$&(93.751) & 0.676 \\
Mobile connectivity                   &  $44.038$&(229.080)& $-97.659$&(71.673)& $141.698$&(240.031)& 0.555 \\
\bottomrule\bottomrule
\end{tabular}
\begin{minipage}{0.93\textwidth}
\footnotesize
\textit{Notes}:  
Each ATE is obtained from a separate stacked difference-in-differences regression with country and year fixed effects and the full set of controls.  
Panel~A compares sale–and–leaseback (SLB) transactions with outright acquisitions.  
Panel~B contrasts deals involving up to 2,000 towers with larger transactions.  
$\Delta$ATE denotes the difference between group-specific effects. Standard errors clustered at the country level appear in parentheses.
\end{minipage}
\end{table}

SLB deals reduce mobile-voice tariffs by roughly 1.1 log point per minute, whereas acquisitions exhibit a (statistically insignificant) increase; the 2 gap is significant at the 1\% level.  No meaningful heterogeneity emerges for data prices or the connectivity metrics.

Transactions covering more than 2,000 towers do not yield larger quality improvements, nor do they translate into stronger price effects; all differences in Panel~B are imprecisely estimated and far from conventional significance thresholds.

\newpage

\section{Conclusion}
\label{sect:conclusion}



This paper combines a newly assembled global database of tower-sharing transactions with a staggered‐adoption difference-in-differences design to measure the effects of large ($\geq$ 1,000-site) tower-sharing transactions on mobile‐telecom outcomes. Across the considered period, the PPP-adjusted price of mobile voice falls by \$1.60 from a 2010 mean of \$3.16, and the price of mobile data declines by \$1.00 from a baseline of \$3.41 per GB. Over the same horizon mobile and internet connections increase substantially, and Internet access expands by 4.7 percentage points in rural households and 3.6 percentage points in women-headed households. Retail market concentration drops by 5.3 percent two to three years after treatment, consistent with intensified downstream rivalry as a transmission mechanism. All estimates are robust to alternative tower-count thresholds, additional controls, and placebo pre-trend tests.

Heterogeneity analysis shows that contractual form dominates deal size. Sale-and-leaseback arrangements in which the incumbent remains an anchor tenant generate the largest price pass-through, whereas increasing a transaction beyond about 2000 sites yields little additional benefit. These patterns suggest that independent tower companies can complement spectrum-allocation and service-based policies aimed at narrowing the digital divide, but that careful attention to contract design is required for the benefits to materialize.

Open questions remain. Future work should examine how passive sharing interacts with active-equipment and spectrum-sharing agreements, quantify the welfare split between consumer surplus and operator profits, and investigate whether access regulation is needed to preserve competitive incentives once tower assets are concentrated in the hands of a few specialised firms. Understanding these margins is critical for designing policies that harness shared infrastructure as an engine of inclusive digital development.

\newpage
\bibliographystyle{authordate1}
\bibliography{9.references}

\newpage
\section*{Appendix}

\makeatother
\setcounter{table}{0}
\setcounter{figure}{0}
\renewcommand\thetable{A-\arabic{table}}
\renewcommand\thefigure{A-\arabic{figure}}

\section{Data and summary statistics}

\begin{table}[htbp]
  \centering
  \caption{List of variables and data sources}
   \scalebox{.7}{
    \begin{tabular}{llll}
    Variable & type  & Label & Source \\
    \hline
    pop   & long  & Total population & WB \\
    cov\_3g & double & 3G network coverage; by population & GSMA \\
    cov\_4g & double & 4G network coverage; by population & GSMA \\
    sim\_subs & double & SIMs per unique mobile subscriber & GSMA \\
    mob\_connect & long  & Total mobile connections & GSMA \\
    mob\_subs & double & Market penetration; unique mobile subscribers & GSMA \\
    int\_connect & long  & Mobile broadband capable connections & GSMA \\
    int\_subs & double & Market penetration; unique mobile internet subscribers & GSMA \\
    hhi   & int   & Herfindahl-Hirschman Index & GSMA \\
    GNIpc\_mdta & double & price of 1.5 GB Mobile broadband, \% of monthly GNI per capita & ITU \\
    GNIpc\_mcell & double & price of Mobile Cellular Low Usage price, \% of monthly GNI per capita & ITU \\
    GNIpc\_mbb & double & price of Mobile Data and Voice Low Usage, \% of monthly GNI per capita & ITU \\
    GNIpc\_mbbh & double & price of Mobile Data and Voice High Usage, \% of monthly GNI per capita & ITU \\
    PPP\_mdta & double & price of 1.5 GB Mobile broadband data, USD PPP & ITU \\
    PPP\_mcell & double & price of Mobile Cellular Low Usage, USD PPP & ITU \\
    PPP\_mbb & double & price of Mobile Data and Voice Low Usage, USD PPP & ITU \\
    PPP\_mbbh & double & price of Mobile Data and Voice High Usage, USD PPP & ITU \\
    USD\_mdta & double & price of 1.5 GB Mobile broadband data, USD & ITU \\
    USD\_mcell & double & price of Mobile Cellular Low Usage, USD & ITU \\
    USD\_mbb & double & price of Mobile Data and Voice Low Usage, USD & ITU \\
    USD\_mbbh & double & price of Mobile Data and Voice High Usage, USD & ITU \\
    int\_indiv & double & \% adults with internet access & Gallup Survey \\
    int\_indiv\_urban & double & \% adults with internet access, urban & Gallup Survey \\
    int\_indiv\_rural & double & \% adults with internet access, rural & Gallup Survey \\
    int\_indiv\_men & double & \% adults with internet access, men & Gallup Survey \\
    int\_indiv\_women & double & \% adults with internet access, women & Gallup Survey \\
    int\_hh & double & \% households with internet access & Gallup Survey \\
    int\_hh\_urban & double & \% households with internet access, urban & Gallup Survey \\
    int\_hh\_rural & double & \% households with internet access, rural & Gallup Survey \\
    int\_hh\_men & double & \% households with internet access, men & Gallup Survey \\
    int\_hh\_women & double & \% households with internet access, women & Gallup Survey \\
    country\_tower\_deals & str26 & country\_tower\_deals & TowerXchange \\
    sites & double & Number of tower sites involved in deal & TowerXchange \\
    dealtype & str11 & Type of deal: acquisition, mll & TowerXchange \\
    deal  & float & dummy variable, 1 if tower deal at year t, 0 if no deal & TowerXchange \\
    ict\_tracker & double & overallscore & ITU \\
    gdp   & double & GDP (current US\$) & WB \\
    gdp\_per\_capita & double & GDP per capita (current US\$) & WB \\
    per\_capita\_ppp & double & GDP per capita, PPP (current international \$) & WB \\
    totaltower & long  & total number of towers & TowerXchange \\
    nbrtowerco & long  & towers owned by towercos & TowerXchange \\
    sharetowerco & double & Share of towerco of total towers & TowerXchange \\
    \hline
    \end{tabular}%
    }
  \label{tab:data}%
\end{table}%

\begin{table}[htbp]
  \centering
  \caption{Number of recorded telecom tower deals}
   \scalebox{.7}{
    \begin{tabular}{lrrrrrrrrrrrrrr}
          & 2008  & 2009  & 2010  & 2011  & 2012  & 2013  & 2014  & 2015  & 2016  & 2017  & 2018  & 2019  & 2020  & \multicolumn{1}{l}{Total} \\
          \hline
    Argentina &       &       &       &       &       &       &       &       & 1     &       &       &       &       & 1 \\
    Bolivia &       &       &       &       &       &       &       &       &       &       &       & 2     &       & 2 \\
    Brazil &       &       &       & 2     & 5     & 9     & 3     & 3     & 1     & 1     &       & 2     & 3     & 29 \\
    Burkina Faso &       &       &       &       &       &       & 1     &       &       &       &       & 1     &       & 2 \\
    Cameroon &       &       &       &       & 1     & 1     &       &       &       &       &       &       &       & 2 \\
    Colombia &       &       &       & 2     &       &       &       &       & 1     & 3     &       & 2     & 1     & 9 \\
    Congo &       &       &       &       &       &       & 1     &       &       &       &       &       &       & 1 \\
    Congo; DR &       &       & 1     &       &       &       &       &       & 1     &       &       &       &       & 2 \\
    Costa Rica &       &       &       &       &       & 1     &       &       &       &       &       &       &       & 1 \\
    Cote d'Ivoire &       &       &       &       & 1     & 1     &       &       &       &       &       &       &       & 2 \\
    Dominican Rep. &       &       &       &       &       &       &       & 1     & 1     &       & 1     &       &       & 3 \\
    Ecuador &       &       &       &       &       &       &       & 1     &       &       &       & 1     &       & 2 \\
    Egypt &       &       &       &       &       &       &       & 1     &       &       &       &       &       & 1 \\
    El Salvador &       &       &       &       &       &       &       &       &       & 1     & 1     &       &       & 2 \\
    Ghana &       &       & 3     &       &       &       & 1     &       &       &       &       & 1     &       & 5 \\
    Guatemala &       &       &       &       &       &       &       &       &       &       & 1     &       &       & 1 \\
    India &       & 3     & 2     &       &       &       &       & 2     & 2     & 2     & 3     &       &       & 14 \\
    Indonesia & 2     &       & 1     & 1     & 4     & 1     & 1     &       & 1     & 1     & 2     & 1     &       & 15 \\
    Jamaica &       &       &       &       &       &       &       &       &       &       & 1     &       &       & 1 \\
    Kenya &       &       &       &       &       &       & 1     &       &       &       & 1     & 1     &       & 3 \\
    Laos  &       &       &       &       &       &       &       &       &       &       &       & 1     &       & 1 \\
    Malawi &       &       &       &       &       &       & 1     &       &       &       &       &       &       & 1 \\
    Malaysia &       &       &       &       &       &       & 1     &       & 2     & 1     &       &       &       & 4 \\
    Mexico &       &       &       & 2     &       & 2     &       &       & 1     & 1     &       & 1     &       & 7 \\
    Myanmar &       &       &       &       &       &       &       & 1     & 1     &       & 1     &       &       & 3 \\
    Nicaragua &       &       &       &       &       &       &       &       & 1     &       &       & 1     &       & 2 \\
    Niger &       &       &       &       &       &       & 1     &       &       &       &       & 1     &       & 2 \\
    Nigeria &       &       & 3     &       &       &       & 3     & 1     & 2     &       &       &       &       & 9 \\
    Pakistan &       &       &       &       &       &       &       &       &       & 2     &       &       &       & 2 \\
    Paraguay &       &       &       &       &       &       &       &       &       & 1     &       &       &       & 1 \\
    Peru  &       &       &       &       & 1     &       &       &       & 1     & 2     &       & 1     & 1     & 6 \\
    Rwanda &       &       &       &       &       &       & 2     &       &       &       &       &       &       & 2 \\
    Senegal &       &       &       &       &       &       &       &       & 1     &       &       &       & 1     & 2 \\
    South Africa &       &       & 1     &       &       &       &       &       & 1     &       &       & 1     &       & 3 \\
    Tanzania &       &       & 1     &       &       & 1     &       &       & 1     &       &       &       &       & 3 \\
    Thailand &       &       &       &       &       &       &       &       &       &       &       & 1     &       & 1 \\
    Uganda &       &       &       & 1     & 2     &       & 1     &       &       &       &       & 1     &       & 5 \\
    Ukraine &       &       &       &       &       &       &       &       & 1     &       &       &       &       & 1 \\
    Vietnam &       &       &       &       &       &       &       &       & 1     &       &       &       &       & 1 \\
    Zambia &       &       &       &       &       &       & 2     &       &       &       &       &       &       & 2 \\
    \hline
    Total & 2     & 3     & 12    & 8     & 14    & 16    & 19    & 10    & 21    & 15    & 11    & 19    & 6     & 156 \\
    \hline
    \end{tabular}%
    }
  \label{tab:deals}%
\end{table}%

\begin{table}[htbp]
  \centering
  \caption{Average number of sites involved in recorded telecom tower deals}
   \scalebox{.7}{
    \begin{tabular}{lrrrrrrrrrrrrrr}
          & 2008  & 2009  & 2010  & 2011  & 2012  & 2013  & 2014  & 2015  & 2016  & 2017  & 2018  & 2019  & 2020  & \multicolumn{1}{l}{Total} \\
          \hline
    Argentina &       &       &       &       &       &       &       &       & 1000  &       &       &       &       & 1000 \\
    Bolivia &       &       &       &       &       &       &       &       &       &       &       & 600   &       & 600 \\
    Brazil &       &       &       & 1012  & 1122  & 1501  & 4250  & 2176  & 1655  & 1200  &       & 807   & 1726  & 1726 \\
    Burkina Faso &       &       &       &       &       &       & 800   &       &       &       &       & 1102  &       & 951 \\
    Cameroon &       &       &       &       & 827   & 1000  &       &       &       &       &       &       &       & 914 \\
    Colombia &       &       &       & 1126  &       &       &       &       & 120   & 483   &       & 589   & 770   & 641 \\
    Congo &       &       &       &       &       &       & 393   &       &       &       &       &       &       & 393 \\
    Congo; DR &       &       & 521   &       &       &       &       &       & 967   &       &       &       &       & 744 \\
    Costa Rica &       &       &       &       &       & 400   &       &       &       &       &       &       &       & 400 \\
    Cote d'Ivoire &       &       &       &       & 931   & 1000  &       &       &       &       &       &       &       & 966 \\
    Dominican Rep. &       &       &       &       &       &       &       & 190   & 545   &       & 1049  &       &       & 595 \\
    Ecuador &       &       &       &       &       &       &       & 130   &       &       &       & 1000  &       & 565 \\
    Egypt &       &       &       &       &       &       &       & 2000  &       &       &       &       &       & 2000 \\
    El Salvador &       &       &       &       &       &       &       &       &       & 202   & 800   &       &       & 501 \\
    Ghana &       &       & 1119  &       &       &       & 900   &       &       &       &       & 1102  &       & 1072 \\
    Guatemala &       &       &       &       &       &       &       &       &       &       & 20    &       &       & 20 \\
    India &       & 6686  & 10975 &       &       &       &       & 21291 & 42790 & 47739 & 21033 &       &       & 23482 \\
    Indonesia & 2118  &       & 1482  & 595   & 854   & 300   & 3500  &       & 2500  & 371   & 1400  & 3100  &       & 1487 \\
    Jamaica &       &       &       &       &       &       &       &       &       &       & 451   &       &       & 451 \\
    Kenya &       &       &       &       &       &       & 981   &       &       &       & 723   & 1102  &       & 935 \\
    Laos  &       &       &       &       &       &       &       &       &       &       &       &       &       &  \\
    Malawi &       &       &       &       &       &       & 219   &       &       &       &       &       &       & 219 \\
    Malaysia &       &       &       &       &       &       & 309   &       &       &       &       &       &       & 309 \\
    Mexico &       &       &       & 1069  &       & 1275  &       &       & 120   & 142   &       & 200   &       & 736 \\
    Myanmar &       &       &       &       &       &       &       & 1250  & 100   &       & 1300  &       &       & 883 \\
    Nicaragua &       &       &       &       &       &       &       &       & 119   &       &       & 150   &       & 135 \\
    Niger &       &       &       &       &       &       & 600   &       &       &       &       & 1102  &       & 851 \\
    Nigeria &       &       & 536   &       &       &       & 5335  & 555   & 648   &       &       &       &       & 2162 \\
    Pakistan &       &       &       &       &       &       &       &       &       & 6850  &       &       &       & 6850 \\
    Paraguay &       &       &       &       &       &       &       &       &       & 1400  &       &       &       & 1400 \\
    Peru  &       &       &       &       & 350   &       &       &       & 900   & 125   &       & 1000  & 760   & 543 \\
    Rwanda &       &       &       &       &       &       & 357   &       &       &       &       &       &       & 357 \\
    Senegal &       &       &       &       &       &       &       &       & 450   &       &       &       & 1220  & 835 \\
    South Africa &       &       & 1400  &       &       &       &       &       & 300   &       &       & 900   &       & 867 \\
    Tanzania &       &       & 1200  &       &       & 1149  &       &       & 185   &       &       &       &       & 845 \\
    Thailand &       &       &       &       &       &       &       &       &       &       &       & 778   &       & 778 \\
    Uganda &       &       &       & 962   & 350   &       & 2681  &       &       &       &       & 1102  &       & 1089 \\
    Ukraine &       &       &       &       &       &       &       &       & 811   &       &       &       &       & 811 \\
    Vietnam &       &       &       &       &       &       &       &       & 1972  &       &       &       &       & 1972 \\
    Zambia &       &       &       &       &       &       & 849   &       &       &       &       &       &       & 849 \\
    \hline
    Total & 2118  & 6686  & 2626  & 996   & 845   & 1244  & 2187  & 5323  & 5190  & 8157  & 6386  & 924   & 1322  & 3276 \\
    \hline
    \end{tabular}%
    }
  \label{tab:sites}%
\end{table}%

\begin{table}[htbp]
  \centering
  \caption{Distribution of deals' size}
   \scalebox{1}{
    \begin{tabular}{lr}
    Percentile & \multicolumn{1}{l}{Number of sites} \\
    \hline
    Min   & 20 \\
    \multicolumn{1}{r}{1\%} & 75 \\
    \multicolumn{1}{r}{5\%} & 119 \\
    \multicolumn{1}{r}{10\%} & 150 \\
    \multicolumn{1}{r}{25\%} & 397 \\
    \multicolumn{1}{r}{50\%} & 916 \\
    \multicolumn{1}{r}{75\%} & 1698 \\
    \multicolumn{1}{r}{90\%} & 4630 \\
    \multicolumn{1}{r}{95\%} & 13000 \\
    \multicolumn{1}{r}{99\%} & 43379 \\
    Max   & 90255 \\
    \hline
    \end{tabular}%
    }
  \label{tab:dealsize}%
\end{table}%

\begin{table}[htbp]
  \centering
  \caption{Tower deals per country, 2008-2021}
   \scalebox{1}{
    \begin{tabular}{lrr}
     Unique deals & \multicolumn{1}{l}{Number  of countries} & \multicolumn{1}{l}{Average sites per deal} \\
    \midrule
    \multicolumn{1}{r}{1} & 10    & 899 \\
    \multicolumn{1}{r}{2} & 12    & 1213 \\
    \multicolumn{1}{r}{3} & 4     & 883 \\
    \multicolumn{1}{r}{4} & 1     & 309 \\
    \multicolumn{1}{r}{5} & 2     & 1157 \\
    \multicolumn{1}{r}{6} & 1     & 627 \\
    \multicolumn{1}{r}{7} & 1     & 561 \\
    \multicolumn{1}{r}{9} & 2     & 1193 \\
    \multicolumn{1}{r}{14} & 1     & 25085 \\
    \multicolumn{1}{r}{15} & 1     & 1622 \\
    \multicolumn{1}{r}{29} & 1     & 1717 \\
    \midrule
    Total & 36    &  \\
    \hline
    \end{tabular}%
    }
  \label{tab:dealscountry}%
\end{table}%

\begin{table}[htbp]
  \centering
  \caption{Shared tower treatment variable}
   \scalebox{1}{
    \begin{tabular}{rrr}
   \multicolumn{1}{l}{Number of treatments} & \multicolumn{1}{l}{Countries} & \multicolumn{1}{l}{Average sites per treatment} \\
    \midrule
    1     & 15    & 1774 \\
    2     & 8     & 1474 \\
    3     & 5     & 3190 \\
    4     & 2     & 12454 \\
    5     & 3     & 4726 \\
    6     & 1     & 328744 \\
    9     & 1     & 50066 \\
    10    & 1     & 22299 \\
    \midrule
    \multicolumn{1}{l}{Total} & 36    &  \\
    \hline
    \end{tabular}%
    }
  \label{tab:towertreatment}%
\end{table}%

\begin{table}[htbp]
  \centering
  \caption{Total and towercos towers}
   \scalebox{.6}{
    \begin{tabular}{lrrrrrrrrrrrrr}
          & \multicolumn{6}{c}{\# towers (mnos \& towercos)} &       & \multicolumn{6}{c}{\%towers (towercos only)} \\
\cmidrule{2-7}\cmidrule{9-14}          & 2015  & 2016  & 2017  & 2018  & 2019  & 2020  &       & 2015  & 2016  & 2017  & 2018  & 2019  & 2020 \\
\hline  
    Afghanistan &       &       &       &       & 6645  & 6917  &       & \multicolumn{1}{l}{} & \multicolumn{1}{l}{} & \multicolumn{1}{l}{} & \multicolumn{1}{l}{} & 0.0\% & 0.0\% \\
    Algeria & 17500 & 17500 & 18000 & \textcolor[rgb]{ 1,  0,  0}{18000} & 19000 & 19350 &       & 0.0\% & 0.0\% & 0.0\% & 0.0\% & 0.0\% & 0.0\% \\
    Angola &       &       & 2500  & 2600  & 3318  & 3318  &       & \multicolumn{1}{l}{} & \multicolumn{1}{l}{} & 0.0\% & 0.0\% & 1.1\% & 1.1\% \\
    Argentina &       & 16000 & 16000 & 16150 & 17252 & 17729 &       & \multicolumn{1}{l}{} & 0.0\% & 0.0\% & 8.3\% & 6.7\% & 9.2\% \\
    Bangladesh & 27000 & 29693 & 30000 & 30000 & 39500 & 33734 &       & 41.9\% & 26.9\% & 27.7\% & 32.7\% & 25.6\% & 28.6\% \\
    Bolivia &       &       &       & 4600  & 4200  & 4490  &       & \multicolumn{1}{l}{} & \multicolumn{1}{l}{} & \multicolumn{1}{l}{} & 0.0\% & 9.5\% & 14.5\% \\
    Brazil & 48606 & 54595 & 55875 & 57127 & 60500 & 64966 &       & 66.2\% & 96.3\% & 94.0\% & 97.9\% & 56.2\% & 70.4\% \\
    Bulgaria &       &       &       &       &       & 8320  &       & \multicolumn{1}{l}{} & \multicolumn{1}{l}{} & \multicolumn{1}{l}{} & \multicolumn{1}{l}{} & \multicolumn{1}{l}{} & 31.5\% \\
    Burkina Faso &       &       &       & 1700  & 2380  & 2517  &       & \multicolumn{1}{l}{} & \multicolumn{1}{l}{} & \multicolumn{1}{l}{} & 41.2\% & 28.2\% & 26.5\% \\
    Cambodia & 9000  & 9250  & 9310  & 9200  & 9200  & 9200  &       & 0.0\% & 0.0\% & 21.5\% & 0.0\% & 39.4\% & 39.4\% \\
    Cameroon &       &       &       & 3200  & 3072  & 3718  &       & \multicolumn{1}{l}{} & \multicolumn{1}{l}{} & \multicolumn{1}{l}{} & 71.4\% & 79.5\% & 65.7\% \\
    Chad  &       &       &       & 2000  & 2000  & 2000  &       & \multicolumn{1}{l}{} & \multicolumn{1}{l}{} & \multicolumn{1}{l}{} & 0.0\% & 0.0\% & 0.0\% \\
    China & 1180000 & 1750000 & 1945384 & 1968000 & 1968000 & 2094464 &       & 100.0\% & 100.0\% & 100.0\% & 100.0\% & 100.0\% & 100.0\% \\
    Colombia &       & 15353 & 15349 & 15553 & 16351 & 4000  &       & \multicolumn{1}{l}{} & 26.2\% & 30.3\% & 32.1\% & 36.9\% & 63.2\% \\
    Congo &       &       &       & 800   & 800   & 848   &       & \multicolumn{1}{l}{} & \multicolumn{1}{l}{} & \multicolumn{1}{l}{} & 48.0\% & 57.9\% & 54.6\% \\
    Congo, Democratic Republic & 4200  & 4350  & 4350  & 4293  & 4293  & 4698  &       & 19.0\% & 41.2\% & 42.2\% & 41.3\% & 53.5\% & 48.9\% \\
    Costa Rica &       & 2924  & 3238  & 3352  & 3889  & 4113  &       & \multicolumn{1}{l}{} & 84.6\% & 86.1\% & 78.1\% & 80.3\% & 50.4\% \\
    Cote d'Ivoire &       &       & 3679  & 4142  & 4271  & 4271  &       & \multicolumn{1}{l}{} & \multicolumn{1}{l}{} & 66.0\% & 60.8\% & 63.7\% & 63.7\% \\
    Egypt & 19000 & 19000 & 19000 & \textcolor[rgb]{ 1,  0,  0}{19000} & 22704 & 24989 &       & 0.0\% & 0.0\% & 0.0\% & 0.0\% & 0.2\% & 0.2\% \\
    El Salvador &       & 1246  & 1267  & 1683  & 1807  & 1811  &       & \multicolumn{1}{l}{} & 19.7\% & 36.9\% & 28.7\% & 41.6\% & 58.9\% \\
    Ethiopia &       &       & 6600  & 6600  & 8000  & 7300  &       & \multicolumn{1}{l}{} & \multicolumn{1}{l}{} & 0.0\% & 0.0\% & 0.0\% & 0.0\% \\
    Gabon &       &       & 1000  & 1000  & 1000  & 1000  &       & \multicolumn{1}{l}{} & \multicolumn{1}{l}{} & 0.0\% & 0.0\% & 0.0\% & 0.0\% \\
    Ghana & 5983  & 5983  & \textcolor[rgb]{ 1,  0,  0}{5983} & 6296  & 6605  & 6609  &       & 71.3\% & 72.4\% & 0.0\% & 76.2\% & 77.3\% & 77.3\% \\
    Guatemala &       & 3593  & 3661  & 3680  & 3908  & 1340  &       & \multicolumn{1}{l}{} & 24.9\% & 26.2\% & 26.6\% & 28.1\% & 70.1\% \\
    Honduras &       & 1200  & 1200  & 1200  & 1200  & 4026  &       & \multicolumn{1}{l}{} & 16.7\% & 16.7\% & 16.7\% & 16.7\% & 29.7\% \\
    India & 450000 & 455521 & 461550 & 461121 & 601800 & 617351 &       & 57.3\% & 75.0\% & 77.0\% & 76.2\% & 84.2\% & 84.0\% \\
    Indonesia & 69458 & 85537 & 93549 & 93378 & 95556 & 98385 &       & 55.3\% & 60.8\% & 64.5\% & 62.0\% & 70.1\% & 65.6\% \\
    Iran  &       & 38000 & 38000 & \textcolor[rgb]{ 1,  0,  0}{38000} & 37106 & 41106 &       & \multicolumn{1}{l}{} & 0.0\% & 0.0\% & 0.0\% & 3.0\% & 12.4\% \\
    Iraq  & 12300 & 12300 & 12300 & \textcolor[rgb]{ 1,  0,  0}{12300} & 14769 & 14769 &       & 0.0\% & 0.0\% & 0.0\% & 0.0\% & 0.0\% & 0.0\% \\
    Jordan & 5900  & 5900  & 5900  & \textcolor[rgb]{ 1,  0,  0}{5900} & 6836  & 6853  &       & 0.0\% & 0.0\% & 0.0\% & 0.0\% & 14.9\% & 14.9\% \\
    Kazakhstan &       &       &       & 15400 & 15400 & 16000 &       & \multicolumn{1}{l}{} & \multicolumn{1}{l}{} & \multicolumn{1}{l}{} & 0.0\% & 0.0\% & 0.6\% \\
    Kenya & 17500 & 6600  & 6600  & 6629  & 7571  & 7661  &       & 0.0\% & 10.6\% & 33.3\% & 29.0\% & 26.6\% & 27.3\% \\
    Laos  &       & 7374  & 7374  & 7374  & 7374  & 7374  &       & \multicolumn{1}{l}{} & 100.0\% & 100.0\% & 100.0\% & 100.0\% & 100.0\% \\
    Lebanon & 2000  & 2000  & 2000  & \textcolor[rgb]{ 1,  0,  0}{2000} & 2600  & 2600  &       & 0.0\% & 0.0\% & 0.0\% & 0.0\% & 0.0\% & 0.0\% \\
    Madagascar &       &       & 2100  & 2020  & 2310  & 2310  &       & \multicolumn{1}{l}{} & \multicolumn{1}{l}{} & 42.9\% & 51.0\% & 51.9\% & 56.3\% \\
    Malawi &       &       &       & 1000  & 1000  & 1000  &       & \multicolumn{1}{l}{} & \multicolumn{1}{l}{} & \multicolumn{1}{l}{} & 100.0\% & 100.0\% & 100.0\% \\
    Malaysia & 20000 & 22117 & 22682 & 22802 & 32412 & 35313 &       & 47.3\% & 54.5\% & 56.2\% & 55.9\% & 71.0\% & 62.0\% \\
    Mexico & 22722 & 27205 & 29159 & 30349 & 32584 & 35242 &       & 98.5\% & 90.1\% & 91.4\% & 90.4\% & 88.3\% & 88.9\% \\
    Mongolia &       &       & 1000  & 1000  & 1000  & 1000  &       & \multicolumn{1}{l}{} & \multicolumn{1}{l}{} & 0.0\% & 0.0\% & 0.0\% & 0.0\% \\
    Morocco & 17000 & 17000 & 17000 & \textcolor[rgb]{ 1,  0,  0}{17000} & 19054 & 21052 &       & 0.0\% & 0.0\% & 0.0\% & 0.0\% & 0.0\% & 0.0\% \\
    Mozambique & 4800  & 4400  & 4400  & 4400  & 4400  & 4400  &       & 0.0\% & 0.0\% & 0.0\% & 100.0\% & 100.0\% & 100.0\% \\
    Myanmar & 7410  & 10750 & 13620 & 15827 & 16000 & 23916 &       & 65.1\% & 66.5\% & 60.4\% & 46.2\% & 73.4\% & 46.6\% \\
    Namibia &       & 2000  & 2000  & 2000  & 749   & 749   &       & \multicolumn{1}{l}{} & 0.0\% & 0.0\% & 37.5\% & 100.0\% & 100.0\% \\
    Nepal &       & 6000  & 6000  & 6000  & 6000  & 6000  &       & \multicolumn{1}{l}{} & 0.0\% & 0.0\% & 0.0\% & 0.0\% & 0.0\% \\
    Nicaragua &       & 1004  & 1115  & 1195  & 1295  & 1810  &       & \multicolumn{1}{l}{} & 65.1\% & 68.6\% & 70.7\% & 71.0\% & 44.2\% \\
    Niger &       &       &       & 1800  & 1800  & 1853  &       & \multicolumn{1}{l}{} & \multicolumn{1}{l}{} & \multicolumn{1}{l}{} & 100.0\% & 100.0\% & 97.1\% \\
    Nigeria & 30941 & 27675 & 28241 & 29652 & 30540 & 31570 &       & 75.1\% & 76.7\% & 81.2\% & 75.7\% & 77.5\% & 78.4\% \\
    Pakistan &       &       &       &       &       & 32000 &       & \multicolumn{1}{l}{} & \multicolumn{1}{l}{} & \multicolumn{1}{l}{} & \multicolumn{1}{l}{} & \multicolumn{1}{l}{} & 6.3\% \\
    Paraguay &       &       & 4250  & 4250  & 4250  & 4296  &       & \multicolumn{1}{l}{} & \multicolumn{1}{l}{} & 58.8\% & 58.8\% & 30.8\% & 33.2\% \\
    Peru  &       & 9118  & 9193  & 10646 & 11202 & 15041 &       & \multicolumn{1}{l}{} & 24.3\% & 9.1\% & 17.6\% & 22.8\% & 42.0\% \\
    Philippines &       & 16300 & 16300 & 17850 & 17850 & 17850 &       & \multicolumn{1}{l}{} & 0.0\% & 0.0\% & 0.0\% & 0.0\% & 0.0\% \\
    Russian Federation &       & 117100 & 117700 & 60850 & 126660 & 140900 &       & \multicolumn{1}{l}{} & 0.0\% & 0.0\% & 0.0\% & 40.3\% & 10.3\% \\
    Rwanda &       &       & 1300  & 1300  & 1300  & 1300  &       & \multicolumn{1}{l}{} & \multicolumn{1}{l}{} & 59.0\% & 62.6\% & 62.6\% & 62.6\% \\
    Senegal & 2900  & 3350  & 3350  & 3151  & 3925  & 4045  &       & 0.0\% & 0.0\% & 0.0\% & 0.0\% & 0.0\% & 30.2\% \\
    Serbia &       &       &       &       & 5146  & 5146  &       & \multicolumn{1}{l}{} & \multicolumn{1}{l}{} & \multicolumn{1}{l}{} & \multicolumn{1}{l}{} & 32.0\% & 32.0\% \\
    South Africa & 22288 & 25000 & 30431 & 28581 & 30183 & 30560 &       & 9.4\% & 9.9\% & 20.6\% & 32.9\% & 31.3\% & 37.3\% \\
    Sri Lanka &       & 7000  & 7500  & 7500  & 8000  & 8000  &       & \multicolumn{1}{l}{} & 0.0\% & 0.0\% & 0.0\% & 1.6\% & 1.6\% \\
    Tanzania &       & 8800  & 7415  & 8278  & 8278  & 8422  &       & \multicolumn{1}{l}{} & 40.7\% & 47.1\% & 42.4\% & 42.5\% & 43.6\% \\
    Thailand & 47483 & 52483 & 52483 & 52483 & 52483 & 52483 &       & 71.6\% & 64.8\% & 64.8\% & 64.8\% & 64.8\% & 64.8\% \\
    Tunisia & 7000  & 7000  & 7000  & \textcolor[rgb]{ 1,  0,  0}{7000} & 8383  & 7955  &       & 0.0\% & 0.0\% & 0.0\% & 0.0\% & 0.0\% & 0.0\% \\
    Turkey &       &       &       &       & 49032 & 50215 &       & \multicolumn{1}{l}{} & \multicolumn{1}{l}{} & \multicolumn{1}{l}{} & \multicolumn{1}{l}{} & 18.7\% & 19.7\% \\
    Uganda & 2547  & 3485  & 3517  & 3554  & 3816  & 4123  &       & 100.0\% & 85.9\% & 85.1\% & 84.2\% & 79.0\% & 80.6\% \\
    Ukraine &       &       & 12000 & 21601 & 21600 & 21655 &       & \multicolumn{1}{l}{} & \multicolumn{1}{l}{} & 10.0\% & 5.6\% & 5.6\% & 5.3\% \\
    Vietnam & 55000 & 70000 & 70000 & 90000 & 90000 & 90000 &       & 0.0\% & 0.0\% & 0.0\% & 0.0\% & 3.7\% & 3.7\% \\
    Zambia &       &       &       & 2300  & 2300  & 3164  &       & \multicolumn{1}{l}{} & \multicolumn{1}{l}{} & \multicolumn{1}{l}{} & 74.5\% & 81.8\% & 59.5\% \\
    Zimbabwe &       & 1400  & 2700  & 2700  & 2700  & 3000  &       & \multicolumn{1}{l}{} & 0.0\% & 0.0\% & 0.0\% & 0.0\% & 0.0\% \\
    \hline
    Average & 81098 & 69375 & 62349 & 53760 & 56174 & 57791 &       & 34.4\% & 31.0\% & 30.2\% & 35.9\% & 38.3\% & 38.1\% \\
    \hline
    \end{tabular}%
    }
  \label{tab:towers}%
\end{table}%

\begin{table}[htbp]
  \centering
  \caption{Deal size in \% of the number of towers}
   \scalebox{.7}{
    \begin{tabular}{lrrrrrrr}
          & 2015  & 2016  & 2017  & 2018  & 2019  & 2020  & \multicolumn{1}{l}{Average} \\
    \midrule
    Argentina & \multicolumn{1}{l}{} & 6.3\% & \multicolumn{1}{l}{} & \multicolumn{1}{l}{} & \multicolumn{1}{l}{} & \multicolumn{1}{l}{} & 6.3\% \\
    Bolivia & \multicolumn{1}{l}{} & \multicolumn{1}{l}{} & \multicolumn{1}{l}{} & \multicolumn{1}{l}{} & 28.6\% & \multicolumn{1}{l}{} & 28.6\% \\
    Brazil & 13.4\% & 3.0\% & 2.1\% & \multicolumn{1}{l}{} & 2.7\% & 8.0\% & 5.8\% \\
    Burkina Faso & \multicolumn{1}{l}{} & \multicolumn{1}{l}{} & \multicolumn{1}{l}{} & \multicolumn{1}{l}{} & 46.3\% & \multicolumn{1}{l}{} & 46.3\% \\
    Colombia & \multicolumn{1}{l}{} & 0.8\% & 9.4\% & \multicolumn{1}{l}{} & 7.2\% & 19.3\% & 9.2\% \\
    Congo; DR & \multicolumn{1}{l}{} & 22.2\% & \multicolumn{1}{l}{} & \multicolumn{1}{l}{} & \multicolumn{1}{l}{} & \multicolumn{1}{l}{} & 22.2\% \\
    Egypt & 10.5\% & \multicolumn{1}{l}{} & \multicolumn{1}{l}{} & \multicolumn{1}{l}{} & \multicolumn{1}{l}{} & \multicolumn{1}{l}{} & 10.5\% \\
    El Salvador & \multicolumn{1}{l}{} & \multicolumn{1}{l}{} & 15.9\% & 47.5\% & \multicolumn{1}{l}{} & \multicolumn{1}{l}{} & 31.7\% \\
    Ghana & \multicolumn{1}{l}{} & \multicolumn{1}{l}{} & \multicolumn{1}{l}{} & \multicolumn{1}{l}{} & 16.7\% & \multicolumn{1}{l}{} & 16.7\% \\
    Guatemala & \multicolumn{1}{l}{} & \multicolumn{1}{l}{} & \multicolumn{1}{l}{} & 0.5\% & \multicolumn{1}{l}{} & \multicolumn{1}{l}{} & 0.5\% \\
    India & 9.5\% & 18.8\% & 20.7\% & 13.7\% & \multicolumn{1}{l}{} & \multicolumn{1}{l}{} & 15.7\% \\
    Indonesia & \multicolumn{1}{l}{} & 2.9\% & 0.4\% & 3.0\% & 3.2\% & \multicolumn{1}{l}{} & 2.4\% \\
    Kenya & \multicolumn{1}{l}{} & \multicolumn{1}{l}{} & \multicolumn{1}{l}{} & 10.9\% & 14.6\% & \multicolumn{1}{l}{} & 12.7\% \\
    Mexico & \multicolumn{1}{l}{} & 0.4\% & 0.5\% & \multicolumn{1}{l}{} & 0.6\% & \multicolumn{1}{l}{} & 0.5\% \\
    Myanmar & 16.9\% & 0.9\% & \multicolumn{1}{l}{} & 8.2\% & \multicolumn{1}{l}{} & \multicolumn{1}{l}{} & 8.7\% \\
    Nicaragua & \multicolumn{1}{l}{} & 11.9\% & \multicolumn{1}{l}{} & \multicolumn{1}{l}{} & 11.6\% & \multicolumn{1}{l}{} & 11.7\% \\
    Niger & \multicolumn{1}{l}{} & \multicolumn{1}{l}{} & \multicolumn{1}{l}{} & \multicolumn{1}{l}{} & 61.2\% & \multicolumn{1}{l}{} & 61.2\% \\
    Nigeria & 1.8\% & 4.7\% & \multicolumn{1}{l}{} & \multicolumn{1}{l}{} & \multicolumn{1}{l}{} & \multicolumn{1}{l}{} & 3.2\% \\
    Paraguay & \multicolumn{1}{l}{} & \multicolumn{1}{l}{} & 32.9\% & \multicolumn{1}{l}{} & \multicolumn{1}{l}{} & \multicolumn{1}{l}{} & 32.9\% \\
    Peru  & \multicolumn{1}{l}{} & 9.9\% & 2.7\% & \multicolumn{1}{l}{} & 8.9\% & 5.1\% & 6.6\% \\
    Senegal & \multicolumn{1}{l}{} & 13.4\% & \multicolumn{1}{l}{} & \multicolumn{1}{l}{} & \multicolumn{1}{l}{} & 30.2\% & 21.8\% \\
    South Africa & \multicolumn{1}{l}{} & 1.2\% & \multicolumn{1}{l}{} & \multicolumn{1}{l}{} & 3.0\% & \multicolumn{1}{l}{} & 2.1\% \\
    Tanzania & \multicolumn{1}{l}{} & 2.1\% & \multicolumn{1}{l}{} & \multicolumn{1}{l}{} & \multicolumn{1}{l}{} & \multicolumn{1}{l}{} & 2.1\% \\
    Thailand & \multicolumn{1}{l}{} & \multicolumn{1}{l}{} & \multicolumn{1}{l}{} & \multicolumn{1}{l}{} & 1.5\% & \multicolumn{1}{l}{} & 1.5\% \\
    Uganda & \multicolumn{1}{l}{} & \multicolumn{1}{l}{} & \multicolumn{1}{l}{} & \multicolumn{1}{l}{} & 28.9\% & \multicolumn{1}{l}{} & 28.9\% \\
    Vietnam & \multicolumn{1}{l}{} & 2.8\% & \multicolumn{1}{l}{} & \multicolumn{1}{l}{} & \multicolumn{1}{l}{} & \multicolumn{1}{l}{} & 2.8\% \\
    \midrule
    Average & 10.4\% & 6.8\% & 10.6\% & 14.0\% & 15.9\% & 15.6\% & 11.9\% \\
    \bottomrule
    \end{tabular}%
    }
  \label{tab:dealshare}%
\end{table}%

\subsection{Coverage Outcomes}

\begin{table}[!htbp]\centering
\caption{Average Post–Treatment Effects on Network Coverage}
\label{tab:cov_ate}
\begin{tabular}{@{}lcc@{}}\toprule\toprule
 & \multicolumn{2}{c}{Aggregated ATE} \\
\cmidrule(lr){2-3}
Outcome & (1) No Covariates & (2) With Covariates \\ 
\midrule
3G coverage  & 0.581 & 1.215 \\[-0.3em]
             & (0.075) & (0.346) \\
4G coverage  & 0.271 & 0.172 \\[-0.3em]
             & (0.109) & (0.112) \\
\bottomrule\bottomrule
\end{tabular}
\vspace{0.5em}
\begin{minipage}{0.85\textwidth}
\footnotesize
\textit{Notes}: Each cell reports the aggregated post–treatment effect estimated with the Sun--Abraham interaction–weighted estimator. Robust standard errors clustered at the country level are in parentheses. Specification (1) is estimated without additional covariates; Specification (2) includes the full control set (GDP per capita, population, and contract type).
\end{minipage}
\end{table}

\end{document}